\documentclass[12pt]{article} 

\usepackage{tipa}
\usepackage{amsmath}
\usepackage{amssymb}
\usepackage{authblk}

\usepackage{natbib}
\usepackage{graphicx}
\usepackage{xr}
\newcommand{\vect}[1]{\boldsymbol{#1}}
\newcommand{\trace}{\operatorname{trace}}
\newcommand{\diag}{\operatorname{diag}}
\newcommand{\AS}{\operatorname{AS}}

\graphicspath{
{./figure/}
{./figure/simu/}
{./figure/bivariate/}
{./figure/multivariate/}
}

\begin{document}

\title{Estimation of Multivariate Wrapped Models for Data in Torus}
\author[1]{Anahita Nodehi}
\author[1]{Mousa Golalizadeh}
\author[2]{Mehdi Maadooliat}
\author[3]{Claudio Agostinelli}
\affil[1]{Department of Statistics, Tarbiat Modares University, Tehran, Iran \texttt{a.nodehi@modares.ac.ir},  \texttt{golalizadeh@modares.ac.ir}}
\affil[2]{Department of Statistics, Marquette University, Milwaukee, USA \texttt{mehdi.maadooliat@marquette.edu}}
\affil[3]{Department of Mathematics, University of Trento, Trento, Italy \texttt{claudio.agostinelli@unitn.it}}

\date{\today}

\maketitle

\begin{abstract}
Multivariate circular observations, i.e. points on a torus are nowadays very common. Multivariate wrapped models are often appropriate to describe data points scattered on $p-$dimensional torus. However, the statistical inference based on this model is quite complicated since each contribution in the log-likelihood involve an infinite sum of indices in $\mathbb{Z}^p$ where $p$ is the dimension of the problem. To overcome this, two estimation procedures based on Expectation-Maximization and Classification Expectation-Maximization algorithms are proposed that works well in moderate dimension size. The performance of the introduced methods are studied by a Monte Carlo simulation and illustrated on three real data sets.   

\noindent \textbf{keyword}: CEM algorithm, EM algorithm, Estimation procedures, Multivariate Wrapped Distributions, Torus.
\end{abstract}

\section{Introduction}
\label{sec:introduction}
There are many problems in applied sciences where a quantity of interest is measured as a direction. \citet{Mardia1972} is one of the first reference in this field and describing how to deal with this kind of data in many subjects. As a simple example of directional data, one can consider one vector with unit length. Clearly, such variable can be specified by an angle on a unit circle provided an initial direction and orientation of the circle have been chosen. This type of data are often refers as circular data as well. One aspect of directional data is that they cannot analyzed using standard methods/models developed for data in the Euclidean space. In other words, one encounters with different topological properties in compare to common Euclidean space. \citet{MardiaJupp2000} and \citet{JammalamadakaSenGupta2001} contain theoretical advances about this field of statistics. Another important reference is \citet{Batschelet1981}.

The Wrapped Normal and the Von Mises are two important distributions on the circle, which resemble on circle the Normal distribution on Euclidean space. For instance, the Von Mises distribution belong to the Exponential family and it is a natural circular analogue of the univariate Normal distribution to which it reduces when the variability in the circular variable are small. In the multivariate setting, its conditional distributions are also Von Mises while its marginal distributions are not. The Wrapped Normal is another circular distribution similar to the univariate Normal. It is symmetric and obtained by wrapping a Normal distribution around the circle. It does not belong to the Exponential family, however, the convolution of two Wrapped Normal variables is also Wrapped Normal \citep{JammalamadakaSenGupta2001}. Its conditional and marginal distribution are Wrapped Normal, too. It appears in the central limit theorem on the circle and in connection with Brownian Motion on the circle (see \citep{Stephens1963}).  

The von Mises distribution is perhaps the most famous in literature, and its fame is due to the possibility of calculating the maximum likelihood parameter estimations analytically, but the extension to multivariate case is still an open problem. In last recent works, \citet{Mardiaetal2007},  \citet{Mardia2008} and \citet{Mardia2010} introduced applications of bivariate and trivariate von Mises distribution. Furthermore, \citet{MardiaVoss2014} studied some properties of a Multivariate Von Mises distribution but its inference requires quite complex estimation procedure. These are some of the reasons of why the Wrapped Normal is often preferred to the Von Mises in the multivariate setting. Moreover, in most of the wrapped distributions there is a correspondence between the in line parameters and wrapped parameters. This allows an easy interpretability of circular parameters and, then, of the inference results.

The Bivariate Wrapped Normal distribution is proposed by \citet{JohnsonWehrly1978} while Multivariate Wrapped Normal distribution is presented in \citet{Baba1981}. Estimation of the Wrapped Normal parameters even in univariate case leads to tough numerical solution since it involves an infinite series. This is one of the main reasons in which some authors, e.g. \citet{Fisher1987b} and \citet{Breckling1989} proposed to approximate this distribution by the Von Mises distribution. \citet{Kent1978} showed that, the von Mises and Wrapped Normal distributions can be well approximated by one another. \citet{Agostinelli2007} proposed an Iterative Reweighting Maximum Likelihood Estimating Equations algorithm (MLE) for univariate Wrapped Normal which is available in the R package \texttt{circular} \citep{AgostinelliLund2017}. \citet{Fisher1994} used the Expectation-Maximization (EM) algorithm techniques to obtain parameter estimates from the Wrapped Normal distribution for AR model with low order. The E-step involves ratios of large infinite sums which need to be approximated at each step making the algorithm computationally inefficient. Moreover, \citet{Coles1998}, \citet{Ravindran2011} and \citet{Ferrari2009} adopted a data augmentation approach to estimate the missing unobserved wrapping coefficients and the other parameters in a Bayesian framework.

As mentioned before, the main difficulty in working with Wrapped Normal is that the form of the density function is constituted by large sums, and cannot be simplified as close form. The likelihood-based inference for such distribution can be very complicated and computationally intensive. We focus on  estimating parameters for both univariate and multivariate Wrapped Normal distribution. Particular, we proposed two innovative algorithms to obtain parameter estimation.

The reminder of this paper is organized as follows. Section \ref{sec:mwd} describes the multivariate wrapped model. Section \ref{sec:estimation} introduces two new algorithms based on Expectation--Maximization and Classification Expectation--Maximization methods for the estimation of the parameters when dealing with the wrapped multivariate normal model. The methods can be extended easily to other multivariate wrapped models. It also includes a description of a method to obtain initial values and a way to extend the proposed estimation methods for the mixed case of torus and linear observations. Section \ref{sec:examples} provides two illustrative examples based on real datasets, while Section \ref{sec:simulations} reports the results of an extensive Monte Carlo experiment. Section \ref{sec:conclusions} gives final comments and remarks. A third example, further results on the two real datasets and complete Monte Carlo summaries can be found in the Supplementary Material.

\section{Multivariate Wrapped Normal}
\label{sec:mwd}

The wrapping approach consists on wrapping a known distribution in the line around a circumference of a circle with a unit radius. A rich class of distributions on the circle can be obtained using the wrapping technique. The procedure is as follows: given a r.v. $X$ in the line then $Y = X \mod 2\pi$ is a r.v. in the circle, accumulating probability over all points $X = (Y + 2 \pi j)$ where $j \in \mathbb{Z}$. If $G$ represents the distribution function on the line, the resulting wrapped distribution $F$ on the circle is given by
\begin{equation*}
F(y)= \sum_{j=-\infty}^{+\infty} [G(y+2 \pi j)- G(2 \pi j) ] \ , \qquad 0 \leq y \leq 2 \pi \ .
\end{equation*}
In particular, if $Y$ has a circular density function $f$ and $g$ is the density function of $X$ then
\begin{equation*}
f(y)= \sum_{j=-\infty}^{+\infty} g(y+2 \pi j) \ , \qquad 0 \leq y \leq 2 \pi \ .
\end{equation*}
By this technique, both discrete and continuous wrapped distributions may be constructed \citep{MardiaJupp2000}. Among the continuous wrapped distributions, Wrapped Normal and Wrapped Cauchy play an important role in data analysis. A Wrapped Normal distribution ($WN(\mu, \sigma^2)$) is obtained by wrapping a $N(\mu,\sigma^2)$ distribution around the circle. This distribution is unimodal and symmetric about the mean $\mu$, the mean resultant length is $\rho= \exp[-\sigma^2/2]$, and as $\rho \rightarrow 0$ the distribution converges to the uniform distribution while as $\rho \rightarrow 1$, it tends to a point mass distribution at $\mu$.

Here we concentrate on the multivariate Wrapped Normal distribution which is obtained by wrapping component-wise a $p$-variate Normal distribution on $p$-variate torus. Similar results and algorithms can be obtained for other multivariate wrapped models.

The bivariate Wrapped Normal distribution is proposed by \citet{JohnsonWehrly1978} and multivariate Wrapped Normal distribution is presented in \citet{Baba1981}. Evaluation of the Wrapped Normal density function can appear difficult even in univariate case because it involves an infinite series. This is one of the main reasons that some authors \citep[e.g.][]{Fisher1987a, Breckling1989, Mardia2008} use the multivariate Von Mises distribution, instead.

\subsection{Equivariance in Wrapper Normal models}

Before move to the estimation problem, discussed in the next Section, we would like to point out that the Wrapper Normal model does not enjoy a full equivariance. Let $\vect{X}$ be a $p$-variate random vector, and given $p$-vector $\vect{b}$ and a full rank $p\times p$ matrix $A$ consider the affine transformation $\vect{W} = A \vect{X} + \vect{b}$. A location estimate $T$ is affine equivariant if $T(\vect{W}) = A T(\vect{X}) + \vect{b}$, while a scatter estimate $S$ is affine equivariant if $S(\vect{W}) = A S(\vect{X}) A^\top$.
Let $\vect{X} \sim N_p(\vect{\mu}_X, \Sigma_X)$, $\vect{b}$ so that $\vect{W} \sim N_p(\vect{\mu}_W, \Sigma_W)$ where $\vect{\mu}_W = A \vect{\mu}_X + \vect{b}$ and $\Sigma_W = A \Sigma_X A^\top$. Define $\vect{U} = \vect{X} \mod 2\pi$ and $\vect{V} = \vect{W} \mod 2\pi$ as two multivariate wrappped normal models on the $p$-torus. In Section SM-1 of the supplementary material we show that the likelihood $L(\mu_W,\Sigma_W | \vect{v}_1, \cdots, \vect{v}_n)$ of the parameters $\mu_W$, $\Sigma_W$ based on the samples $\vect{v}_1, \cdots, \vect{v}_n$ is proportional to $L(\mu_X,\Sigma_X | \vect{u}_1^\ast, \cdots, \vect{u}_n^\ast)$ where $\vect{u}_i^\ast$ is a sample from $\vect{U}^\ast = \vect{X} \mod (2\pi A^{-1} \vect{j})$ and $\vect{j}$ is a $p$-vector of ones and hence it is not proportional to $L(\mu_X,\Sigma_X | \vect{u}_1, \cdots, \vect{u}_n)$.
This fact shows that MLE estimates are not affine equivariant for the multivariate wrapped normal model.

\section{Parameters estimation}
\label{sec:estimation}

Let $\vect{y}_1, \ldots, \vect{y}_n$ be a i.i.d. sample from a Wrapped Normal model $\vect{Y} \sim WN_p(\vect{\mu}, \Sigma)$ in the $p$-torus with mean vector $\vect{\mu}$ and variance-covariance matrix $\Sigma$. We can think of $\vect{y}_i = \vect{x}_i \mod 2\pi$ where $\vect{x}_i$ is a sample from $\vect{X}_i \sim N_p(\vect{\mu}, \Sigma)$.

The log-likelihood of the unknown parameters $\Omega = (\vect{\mu}, \Sigma)$ of a multivariate wrapped normal model is represented by 
\begin{equation} \label{equ:loglik}
\ell(\Omega; \vect{y}_1, \ldots, \vect{y}_n) = \sum_{i=1}^n \log \left[ \sum_{\vect{j}_i \in \mathbb{Z}^p} \phi(\vect{y}_i + 2 \pi \vect{j}_i; \Omega) \right] \,
\end{equation} 
where $\phi$ is the multivariate normal density in $\mathbb{R}^p$ and $\vect{j}_i$ is a vector of indices in $\mathbb{Z}^p$. For the univariate case \citet{Agostinelli2007} proposed an Iteratively Re-Weighting Least Square algorithm which maximizes the log-likelihood. The details of this method are in \citet{Agostinelli2007} and in the function \texttt{mle.wrappednormal} available in R package \texttt{circular} \citep{AgostinelliLund2017}.  For the multivariate case ($p > 1$) a similar algorithm seems to be unfeasible and different technique should be considered. Direct maximization of the log-likelihood is somehow possible for moderate small dimesion, says $p \le 5$, provided a good reparametrization of the parameter space is used. For instance, while dealing with multivariate wrapped normal models, the variance-covariance matrix $\Sigma$ can be reparametrized as described in \citet{PinheiroBates1996}. In our implementationi the Log-Cholesky parameterization, which allows for uncostrained optimization while ensuring positive definite estimate of $\Sigma$, is used. Let $\vect{\sigma}$ be the set of the $p (p+1)/2$ parameters needed to represent $\Sigma$ uniquely and let $\Sigma(\vect{\sigma}) = R(\vect{\sigma})^\top R(\vect{\sigma})$ be the Cholesky decomposition, where $R(\vect{\sigma})$ is a full rank $p\times p$ upper triangular matrix which depends on $\vect{\sigma}$. To ensure that the diagonal elements of $\Sigma$ are positive, the logarithms of the diagonal elements of $R$ is used. In the next two subsections we are going to describe algorithms based on Expectation-Maximization (EM) method \citep{Dempster1977}. \citet{Fisher1994} used the EM algorithm to obtain parameter estimates for (low order) autoregressive models with wrapping normal distributions. However their procedure presents high computational complexity: the E-step involves ratios of large multivariate infinite sums which need to be approximated at each step making the algorithm computationally inefficient. \citet{Coles1998}, \citet{Ravindran2011} and \citet{Ferrari2009} adopted a data augmentation approach to estimate the missing unobserved wrapping coefficients in a Bayesian framework.

\subsection{EM algorithm}
\label{sec:em}

Instead of the log-likelihood in Equation \eqref{equ:loglik} the EM algorithm works with the complete log-likelihood function given by 
\begin{equation} \label{equ:completeloglik}
\ell_C(\Omega; \vect{y}_1, \ldots, \vect{y}_n) = \sum_{i=1}^n \log \left[ \sum_{\vect{j}_i \in \mathbb{Z}^p} v_{i\vect{j}_i} \phi(\vect{y}_i + 2 \pi \vect{j}_i; \Omega) \right] \ .
\end{equation} 
where $v_{i\vect{j}_i}$ is an indicator of the $i$th unit having the $\vect{j}_i$ vector as wrapping coefficients. The algorithm alternates between two steps: Expectation (E) and Maximization (M).
\begin{itemize}
\item {\textbf{E} step:} Compute the (posterior) expectation of the complete log-likelihood by setting $v_{i\vect{j}_i}$ equals to the posterior probability that $\vect{y}_i$ have $\vect{j}_i$ as wrapping coefficients, i.e.
\begin{equation*}
v_{i\vect{j}_i} =  \frac{\phi(\vect{y}_i + 2 \pi \vect{j}_i; \Omega)}{\sum_{\vect{h}_i \in \mathbb{Z}^p} \phi(\vect{y}_i + 2 \pi \vect{h}_i; \Omega)} \ , \qquad \vect{j}_i \in \mathbb{Z}^p, \quad i=1,\ldots,n \ ;
\end{equation*}
\item {\textbf{M} step:} Compute the updated estimates of $\Omega$ by maximizing the complete log-likelihood conditionally on $v_{i\vect{j}_i}$, $i=1,\ldots,n$.
\end{itemize}
At the implementation stage $\mathbb{Z}^p$ is replaced by the Cartesian product $\times_{s=1}^p \mathcal{J}$ where $\mathcal{J} = (-J, -J+1, \ldots, 0, \ldots, J-1, J)$ for some large enough $J$.

Since the M step still involves a complicated maximization problem, we introduce a modification based on the variance decomposition formula.  We call this final algorithm the Variance Decomposition EM algorithm. Fixing $J$, at stage $k$ of the algorithm, let $\vect{\mu}^{(k)}$ and $\Sigma^{(k)}$ the estimates, for $i$th observation we are going to recompute $\vect{y}_i$ ($i=1, \cdots, n$) such that each components of $\vect{y}_i - \vect{\mu}^{(k)}$ are expressed in the interval $(-\pi, \pi]$, this prevents the use of large values of $J$ in order to have a good approximation. We build a data matrix $\tilde{Y}_i$ of dimension $(2 \times J+1)^p \times p$ with row entries on the form
\begin{equation*}
\tilde{\vect{y}}_r = \vect{y}_i + 2 \pi \vect{J}_r = (y_{i1} + 2 \pi j_{r1}, \ y_{i2} + 2 \pi j_{r2}, \ \ldots, \ y_{ip} + 2 \pi j_{rp}) \qquad r=1, \ldots, (2\times J+1)^p
\end{equation*}
where the vector $\vect{J}_r = (j_{r1}, \cdots, j_{rp})$ is one of the $(2 \times J+1)^p$ rows of the matrix obtained by the Cartesian product $\times_{s=1}^p \mathcal{J}$. Let $\tilde{\vect{w}}_i$ be a weights vector with entries $\tilde{w}_{r} = \phi(\tilde{\vect{y}}_{r}, \vect{\mu}^{(k)}, \Sigma^{(k)})$ where $\phi(\cdot, \vect{\mu}, \Sigma)$ is the density of the $p$-multivariate normal distribution with mean vector $\vect{\mu}$, and covariance matrix $\Sigma$.

Let $\tilde{\vect{\mu}}_{i}$ and $\tilde{\Sigma}_{i}$ be the weighted sample mean and weighted sample covariance based on the $\tilde{Y}_i$ data and weights $\tilde{\vect{w}}_i$. Let $M$ be the matrix with row entries $\tilde{\vect{\mu}}_{i}$ and $C$ the sample covariance of $M$ data. Then, we update the parameters as follows
\begin{align*}
\vect{\mu}^{(k+1)} & = \frac{1}{n} \sum_{i=1}^n \tilde{\vect{\mu}}_{i} \\
\Sigma^{(k+1)} & = \frac{1}{n} \sum_{i=1}^n \tilde{\Sigma}_{i} + C \ ,
\end{align*}
where $\tilde{\vect{\mu}}_{i}$ and $\tilde{\Sigma}_{i}$ are the conditional means and conditional (within) variance matrix respectively, while $C$ is the between variance matrix. This algorithm can be easily implemented in a parallel way. First, computations for each observations can be performed independently at each stage. Second, the computation of each of the component of the weights vector $\tilde{w}_i$ can also be performed in parallel. Notice that, at each step, the log-likelihood is not decreasing and hence the algorithm convergences.

\subsection{Classification EM algorithm}
\label{sec:cem}

An alternative algorithm for this estimation problem is the Classifiction EM (CEM) algorithm \citep{Celeux1992} where the E step is followed by a C step (Classification step) in which $v_{i\vect{j}_i}$ is estimated either $0$ or $1$, so that to the $i$th observation is associated the most likely $\vect{j}_i$ vector. In our context this reduces the complete log-likelihood to the following ``classification'' log-likelihood
\begin{equation} \label{equ:classificationloglik}
\ell^C(\Omega, \vect{j}_1, \ldots, \vect{j}_n; \vect{y}_1, \ldots, \vect{y}_n) = \sum_{i=1}^n \log \phi(\vect{y}_i + 2 \pi \vect{j}_i; \Omega) \ .
\end{equation}
in which the $\vect{j}_i \in \mathbb{Z}^p$ ($i=1, \ldots, n$) are treated as unknown parameters. The procedures at stage $k$ is then performed as follows
\begin{itemize}
\item {\textbf{E} step:} Compute as before
\begin{equation*}
v_{i\vect{j}_i} =  \frac{\phi(\vect{y}_i + 2 \pi \vect{j}_i; \Omega)}{\sum_{\vect{h}_i \in \mathbb{Z}^p} \phi(\vect{y}_i + 2 \pi \vect{h}_i; \Omega)} \ , \qquad \vect{j}_i \in \mathbb{Z}^p \quad i=1,\ldots,n \ ;
\end{equation*}
\item {\textbf{C} step:} Let $\hat{\vect{j}_i} = \arg\max_{\vect{h}_i \in \mathbb{Z}^p} v_{i\vect{h}_i}$
\item {\textbf{M} step:} Compute the updated estimates of $\Omega$ by maximizing the classification log-likelihood conditionally on $\hat{\vect{j}}_i$ ($i=1,\ldots,n$).
\end{itemize}
As in the EM algorithm at the implementation stage $\mathbb{Z}^p$ is replaced by the Cartesian product $\times_{s=1}^p \mathcal{J}$ for some large enough $J$. The $\hat{\vect{j}}_i$ plays the role of an offset in the classification log-likelihood and hence the M step is straightforward. Note that, at each stage, the classification algorithm provides also an estimate of the original unobserved sample $\vect{x}_1, \ldots, \vect{x}_n$, obtained as $\hat{\vect{x}}_i = \vect{y}_i + 2\pi \hat{\vect{j}}_i$, ($i=1,\ldots,n$).

\subsection{Extending to torus and linear variables}
\label{sec:extending}
Extension to the mixed case of torus and linear observations can be obtained easily for both the EM and the CEM algorithm. Suppose that 
\begin{equation*}
\vect{X} = 
\begin{bmatrix}
\vect{X}_1 \\ \vect{X}_2 
\end{bmatrix}
\sim N_{p_1+p_2}\left(
\begin{bmatrix}
\vect{\mu}_1 \\ \vect{\mu}_2 
\end{bmatrix}
;
\begin{bmatrix}
\Sigma_{11} & \Sigma_{12} \\ 
\Sigma_{21} & \Sigma_{22} 
\end{bmatrix}
\right).
\end{equation*}
Here, our samples are from the joint vector $(\vect{Y}_1, \vect{X}_2)$ where $\vect{Y}_1 = \vect{X}_1 \mod 2\pi$ that is $\vect{Y}_1 \sim WN_{p_1}(\vect{\mu}_1;\Sigma_{11})$. For the CEM algorithm, we suggest to perfom the algorithm in the previous Section \ref{sec:cem} and obtain the final estimate $\hat{\vect{x}}_{1i}$, ($i=1,\ldots,n$). Consider, for the $i$th observation the whole $p_1+p_2$ vector $(\hat{\vect{x}}_{1i}, \vect{x}_{2i})$ and perform the MLE in order to obtain estimate of the remaining components $\vect{\mu}_2$, $\Sigma_{12}$ and $\Sigma_{22}$. Notice that only the estimate of $\Sigma_{12}$ really need the joint vector. For the EM algorithm, we suggest to perform the algorithm in the previous Section \ref{sec:em} and obtain the final $\tilde{\vect{\mu}}_{1i}$, ($i=1,\ldots,n$). These, can be used as an estimate of the unknown observations $\vect{x}_{1i}$. By considering the whole $p_1+p_2$ vector $(\tilde{\vect{\mu}}_{1i}, \vect{x}_{2i})$ an estimate of $\Sigma_{12}$ can be obtained. The other missed components can be estimate using only the complete samples $\vect{x}_{2i}$.

\subsection{Initial values}
\label{sec:initial}
The introduced algorithms need initial values. We suggest to use the circular means and $-2\log(\hat{\rho}_r)$ where $\hat{\rho}_r$ is the sample mean resultant length for mean vector $\vect{\mu}^{(0)}$ and the variances $\sigma_{rr}^{(0)}$ ($r=1, \ldots, p$), respectively. Following \citep{JammalamadakaSenGupta2001} the circular correlation coefficient between two samples of angles $\vect{x}$ and $\vect{y}$ is defined as
\begin{equation*}
\rho_c(\vect{x}, \vect{y}) = \frac{\sum_{i=1}^n \sin(x_i - \bar{x})\sin(y_i - \bar{y})}{(\sum_{i=1}^n \sin(x_i - \bar{x})^2 \sum_{i=1}^n \sin(y_i - \bar{y})^2)^{1/2}}
\end{equation*}
where $\bar{x}$ and $\bar{y}$ are the circular means. We let $\sigma_{rs}^{(0)} = \rho_c(\vect{y}_r, \vect{y}_s) \sigma_{rr}^{(0)} \sigma_{ss}^{(0)}$ ($r \neq s$) for the covariances initial values, this ensures that the initial matrix $\Sigma^{(0)}$ is a full rank covariance matrix.

\section{Real Data Application}
\label{sec:examples}

We consider two examples. The first example is concerned about a Protein data set which is bivariate. The results of this example are reported in the next Section \ref{sec:protein}. The second example is on RNA and it is analyzed in Section \ref{sec:rna}, in this case observations are on 7-torus, i.e. the 7-dimensional variable lying on torus. A third example related to wind direction is univariate and its analysis is reported in Section SM-2 of the Supplementary Material.

\subsection{Protein data set: bivariate case}
\label{sec:protein}

A collection of data sets called SCOP.1 about protein structure analyzed in \citet{Najibi2017} and described in their Supplementary Material are used as an illustrative example. The collection contains bivariate information about $63$ protein domains that were randomly selected from three remote Protein classes in the Structural Classification of Proteins (SCOP). The class labels are available and we can check the homogeneity in each of three clusters by appling our techniques. 

The constituents of the collection of protein domains are as follows. Cluster 1: 19 domains from ``All beta proteins/ Immunoglobulin-like beta-sandwich /Immunoglobulin/ V set domains (antibody variable domain-like)/ Immunoglobulin light chain kappa variable domain, VL-kappa/ Human (Homo sapiens)''. Sample sizes are in the range between 104 and 106. Cluster 2: 26 domains from ``Alpha and beta proteins (a/b)/TIM beta, alpha-barrel/ Triosephosphate isomerase (TIM)/ Triosephosphate isomerase (TIM)/ Triosephosphate isomerase/ Chicken (Gallus gallus)''. Sample sizes are in the range between 233 and 244. Cluster 3: 18 domains from ``Alpha and beta proteins (a+b)/ Microbial ribonucleases/ Microbial ribonucleases/ Bacterial ribonucleases/ Barnase/ Bacillus amyloliquefaciens''. Sample sizes are in the range between 104 and 107.

\begin{figure}
\begin{center}
\includegraphics[height=0.3\textwidth]{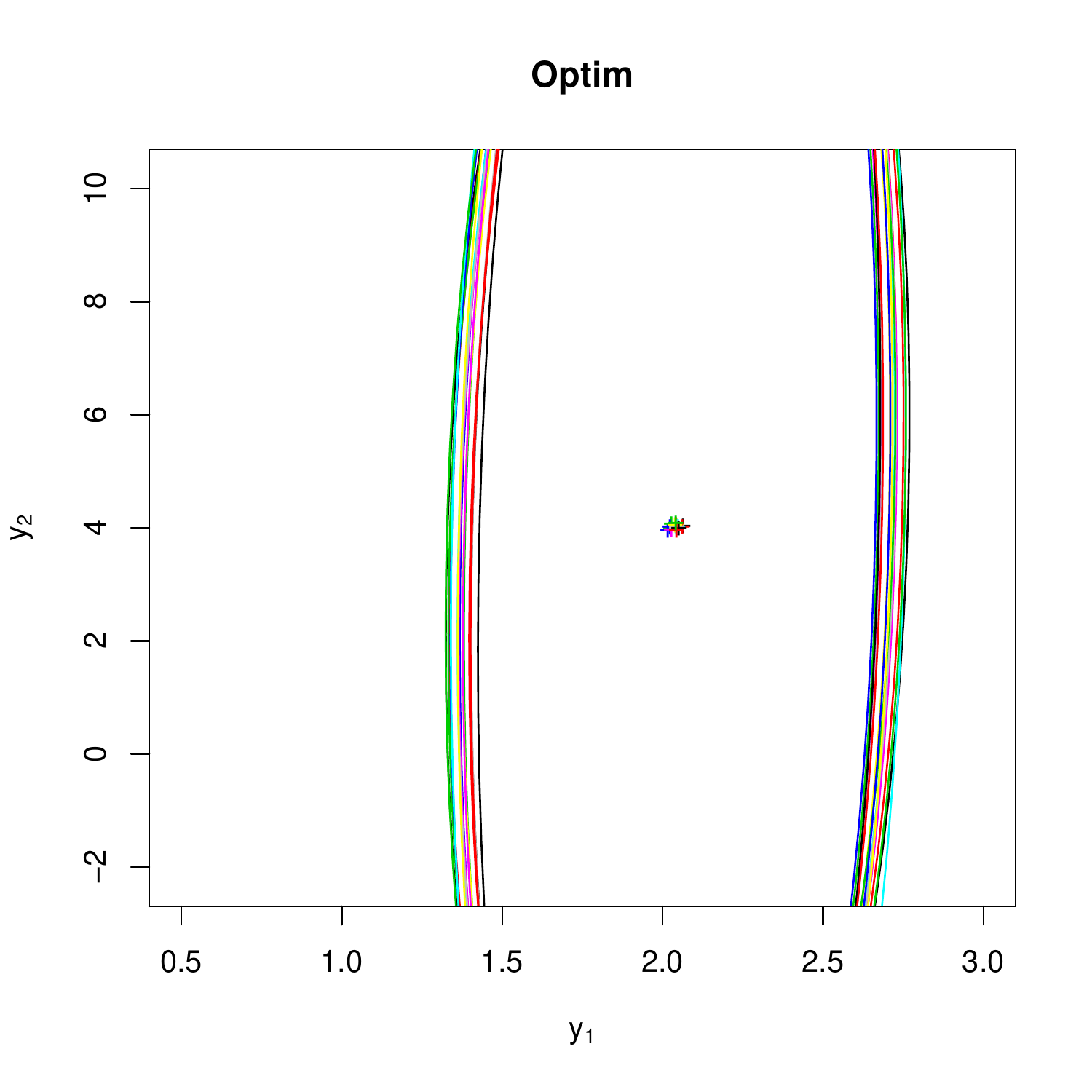}
\includegraphics[height=0.3\textwidth]{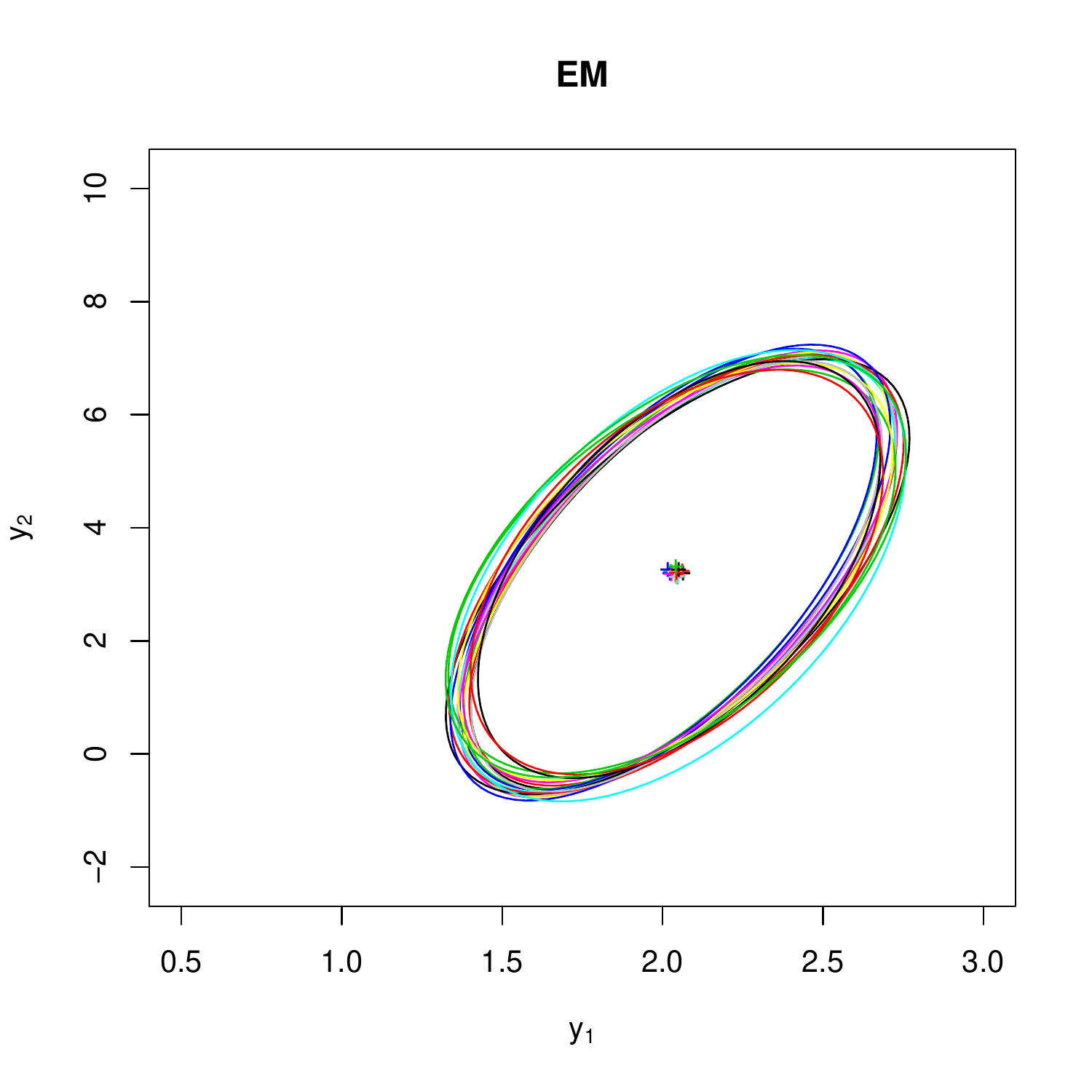}
\includegraphics[height=0.3\textwidth]{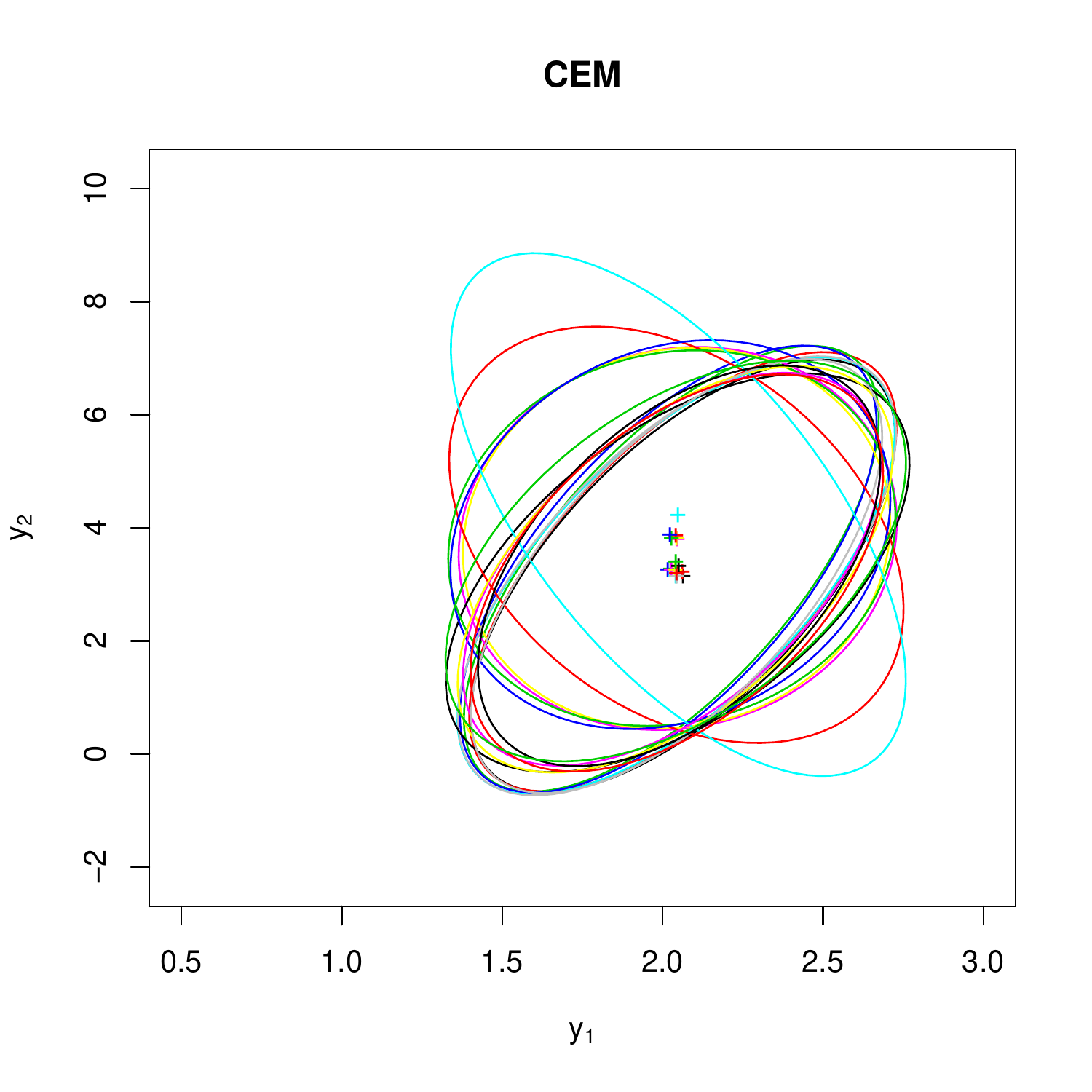} \\
\includegraphics[height=0.3\textwidth]{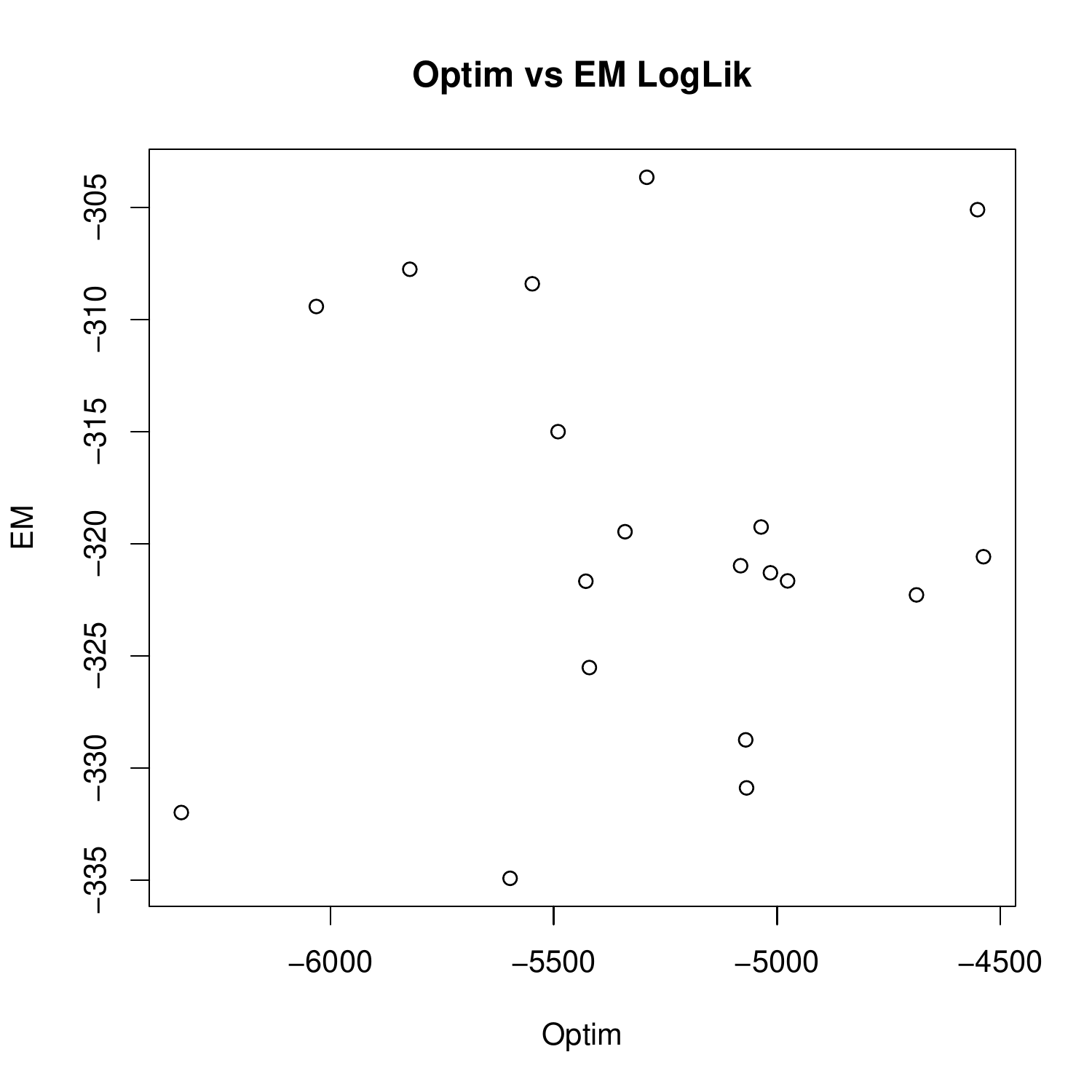}
\includegraphics[height=0.3\textwidth]{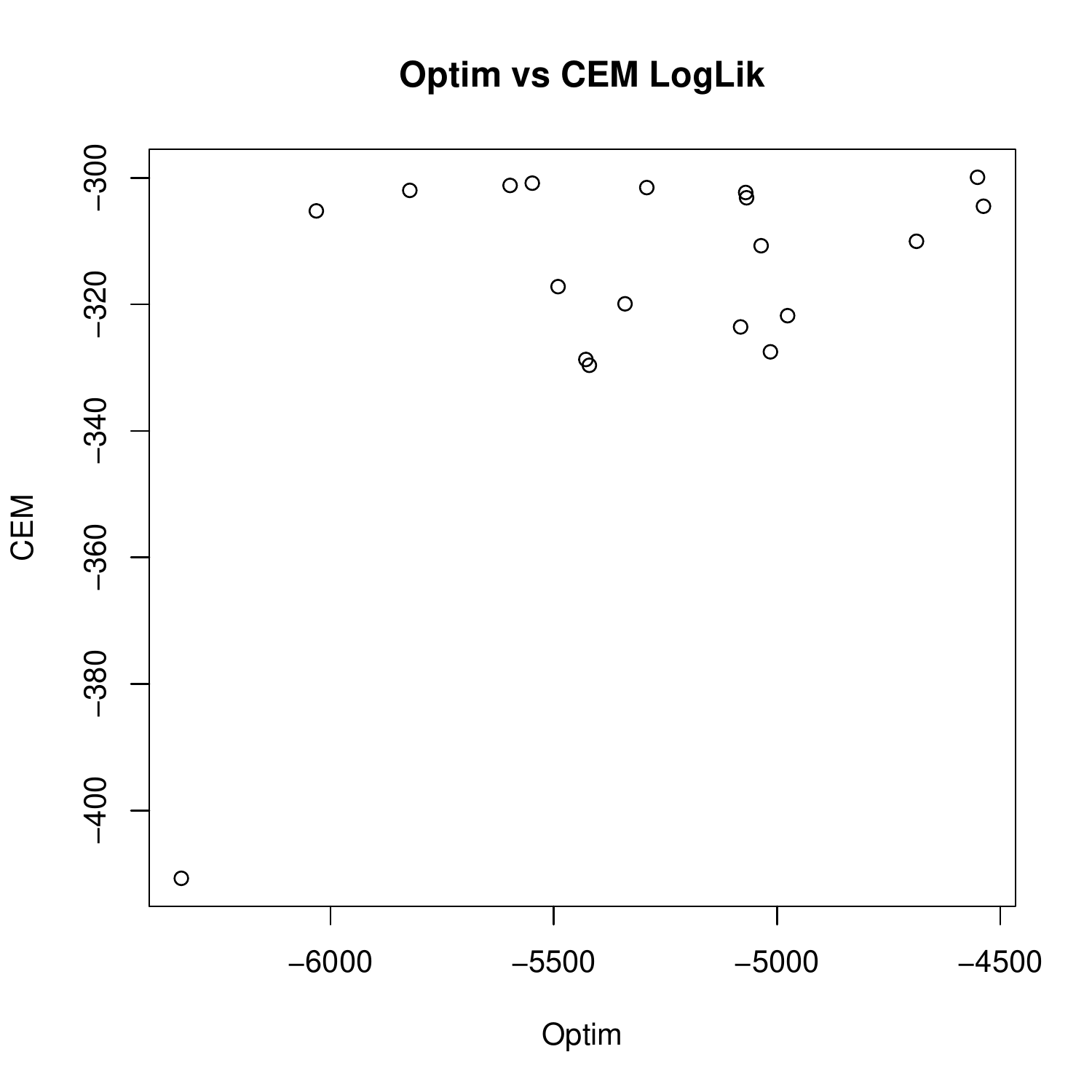}
\includegraphics[height=0.3\textwidth]{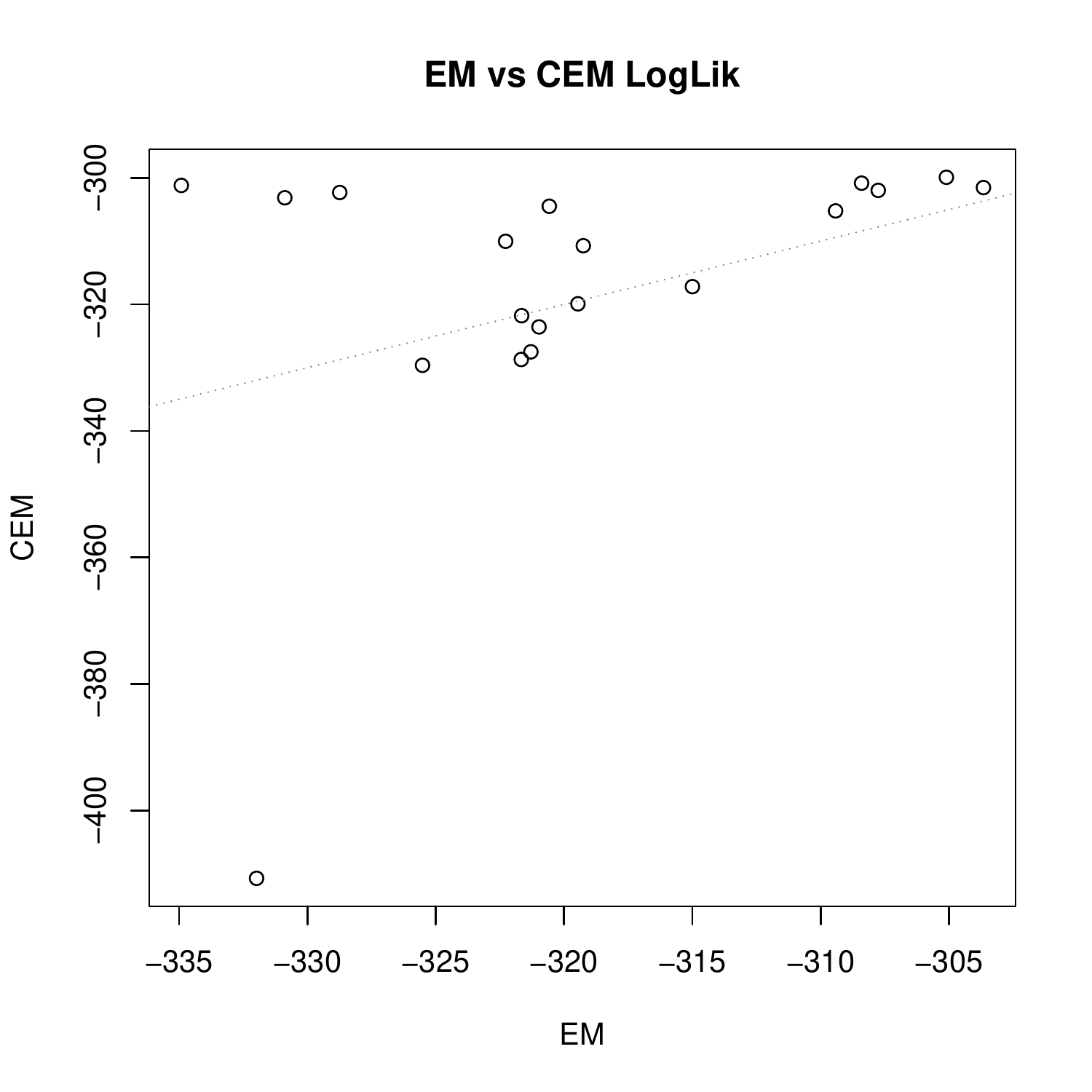}
\end{center}
\caption{Protein data set, cluster 1. First row: estimated means and $95\%$ ellipsoid confidence using optim, EM and CEM. Second row: comparison of the log-likelihood at the estimated values.}
\label{fig:protein:1}
\end{figure}

\begin{figure}
\begin{center}
\includegraphics[height=0.3\textwidth]{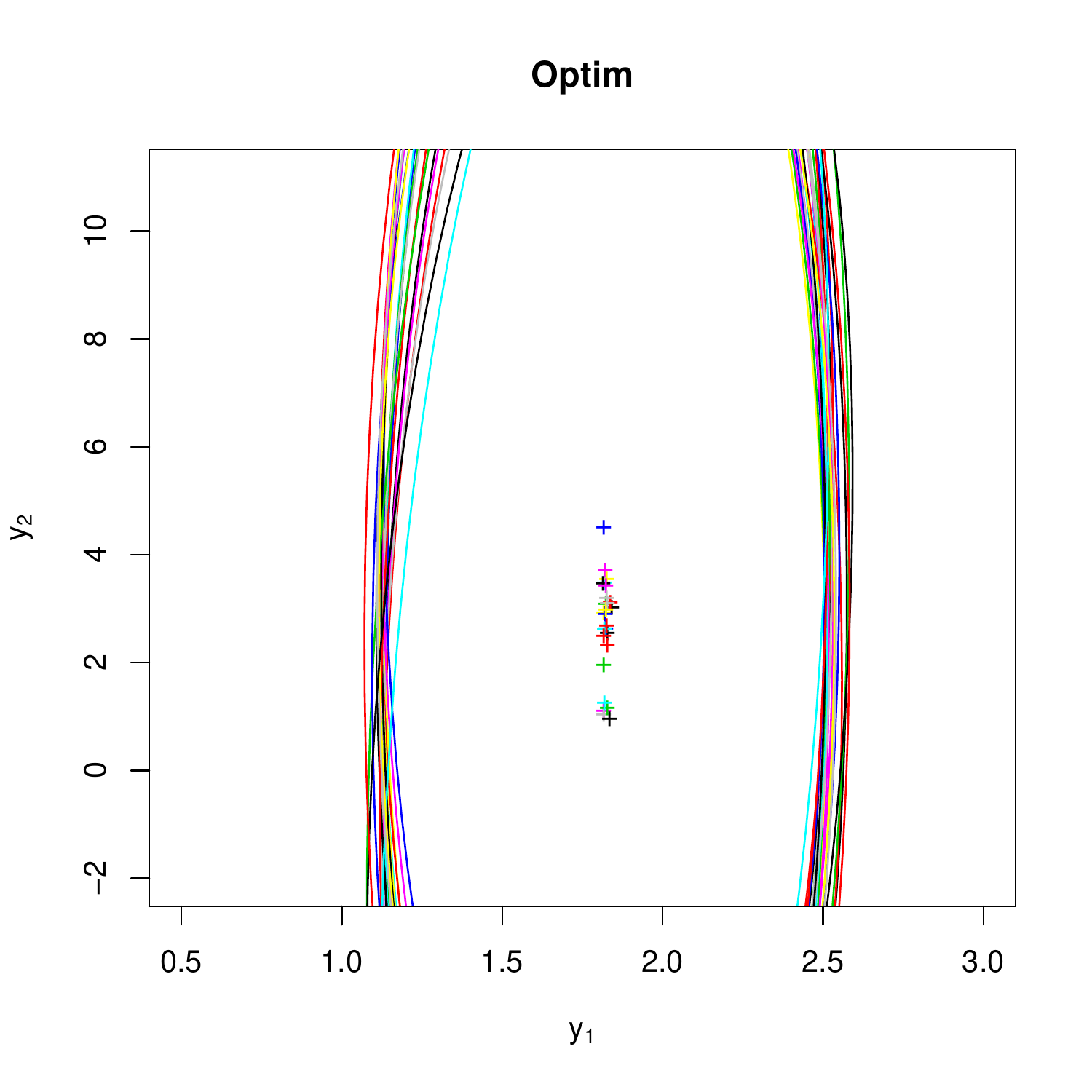}
\includegraphics[height=0.3\textwidth]{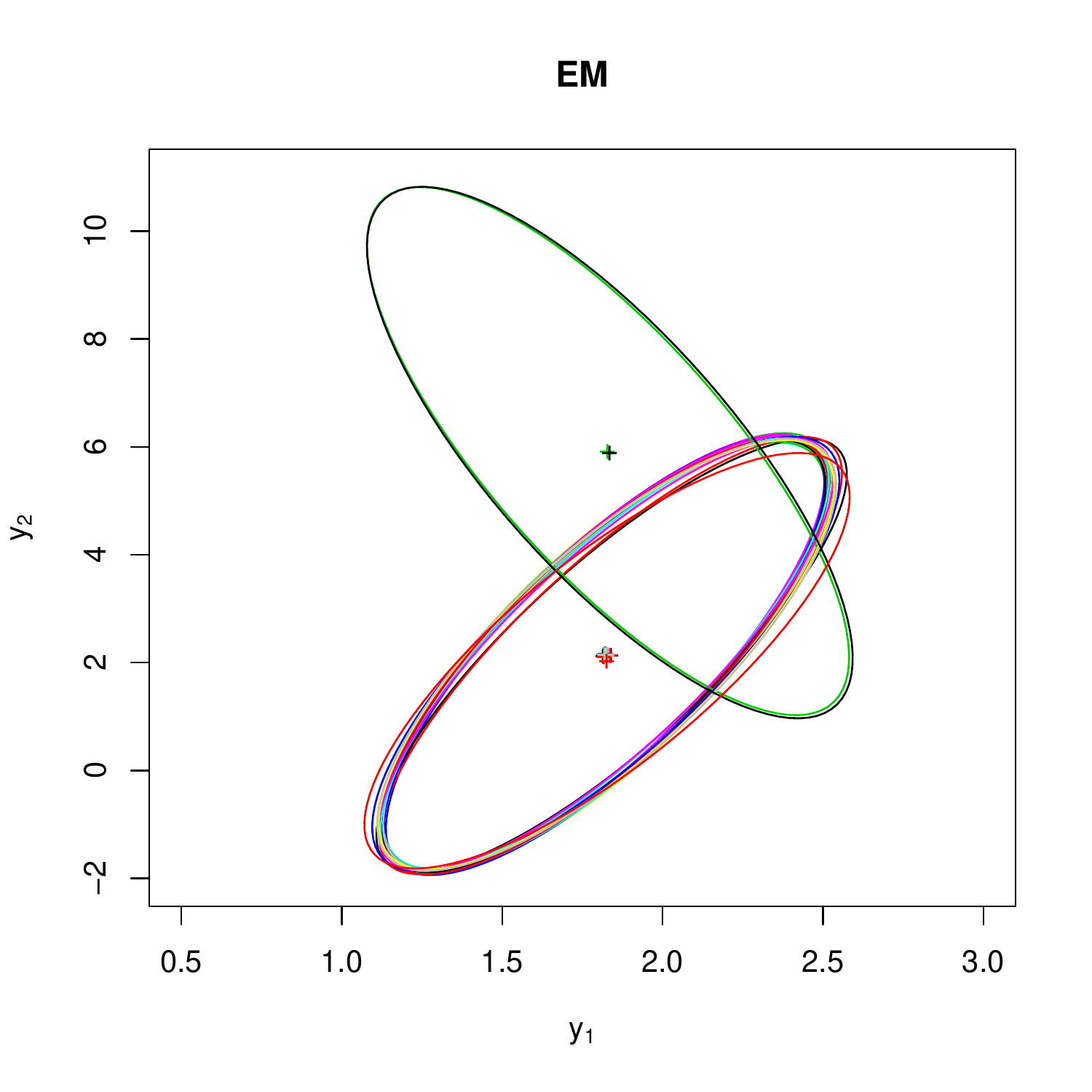} 
\includegraphics[height=0.3\textwidth]{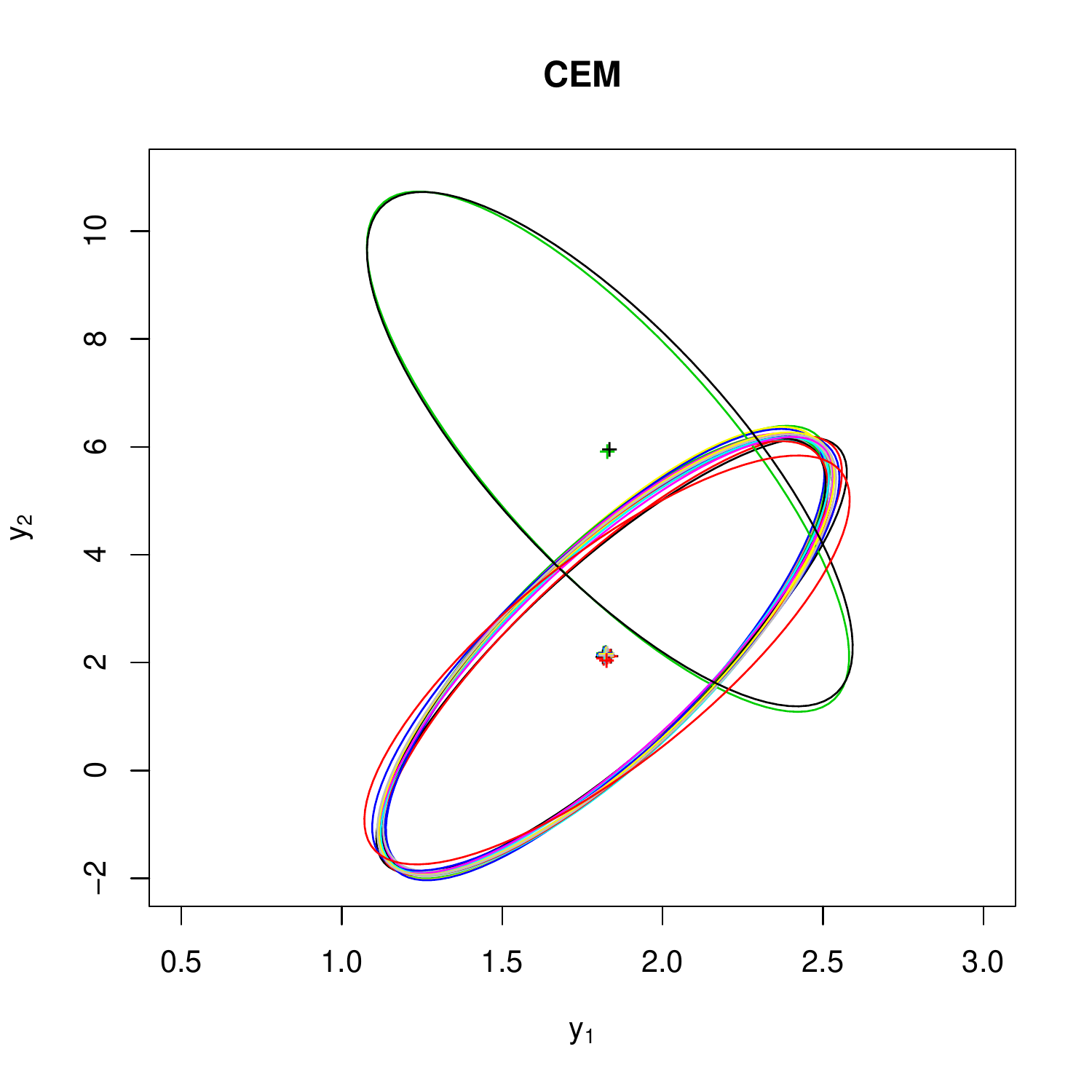} \\
\includegraphics[height=0.3\textwidth]{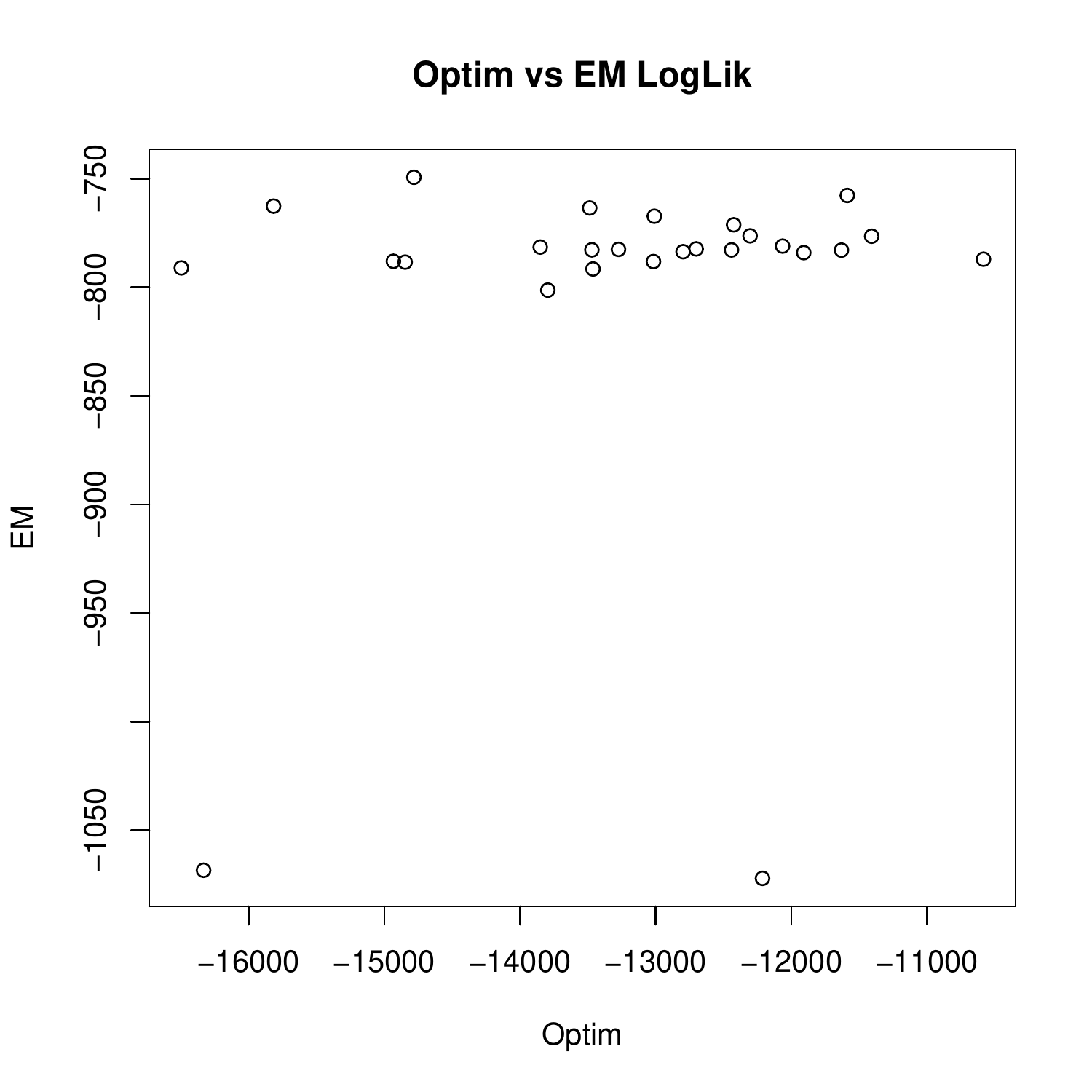}
\includegraphics[height=0.3\textwidth]{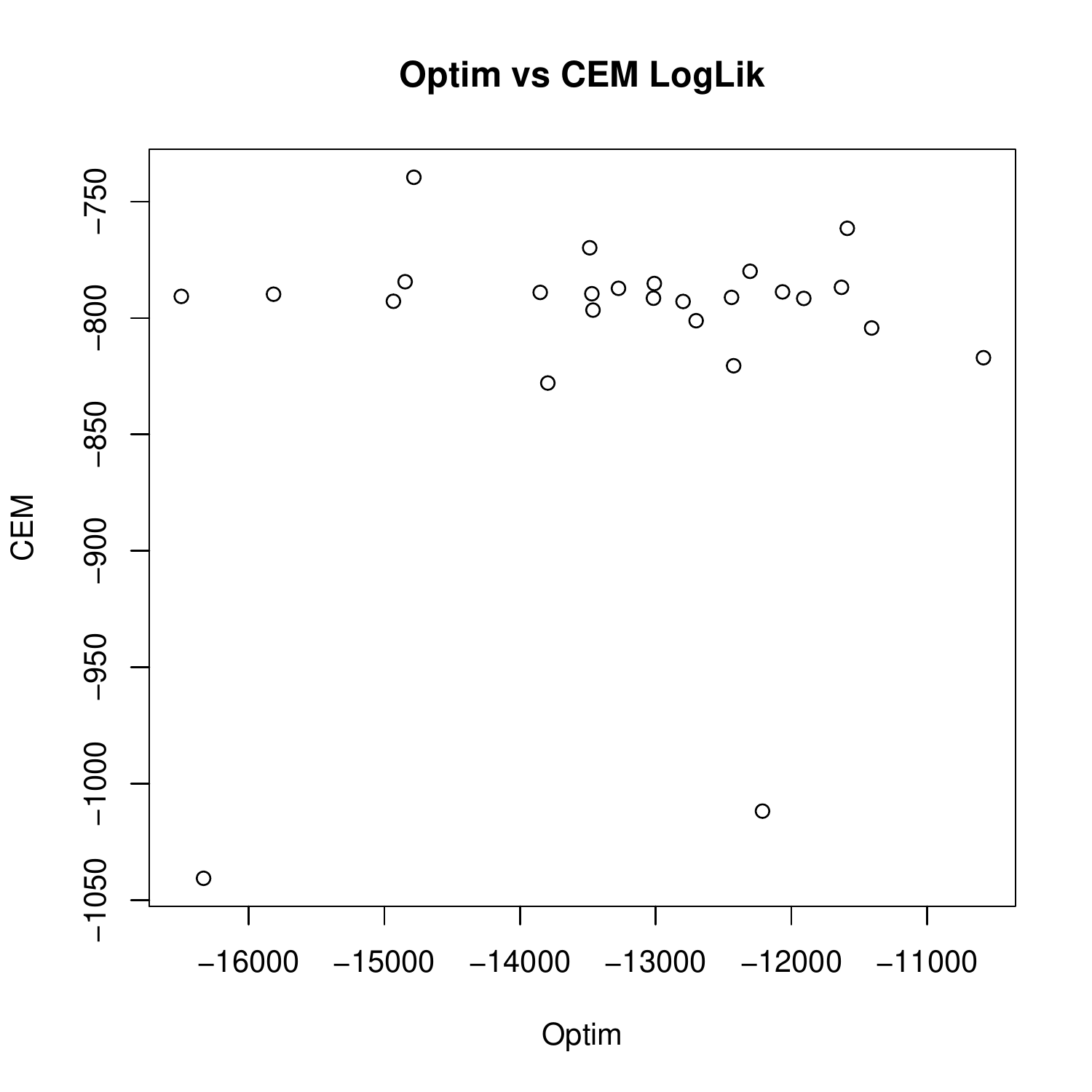}
\includegraphics[height=0.3\textwidth]{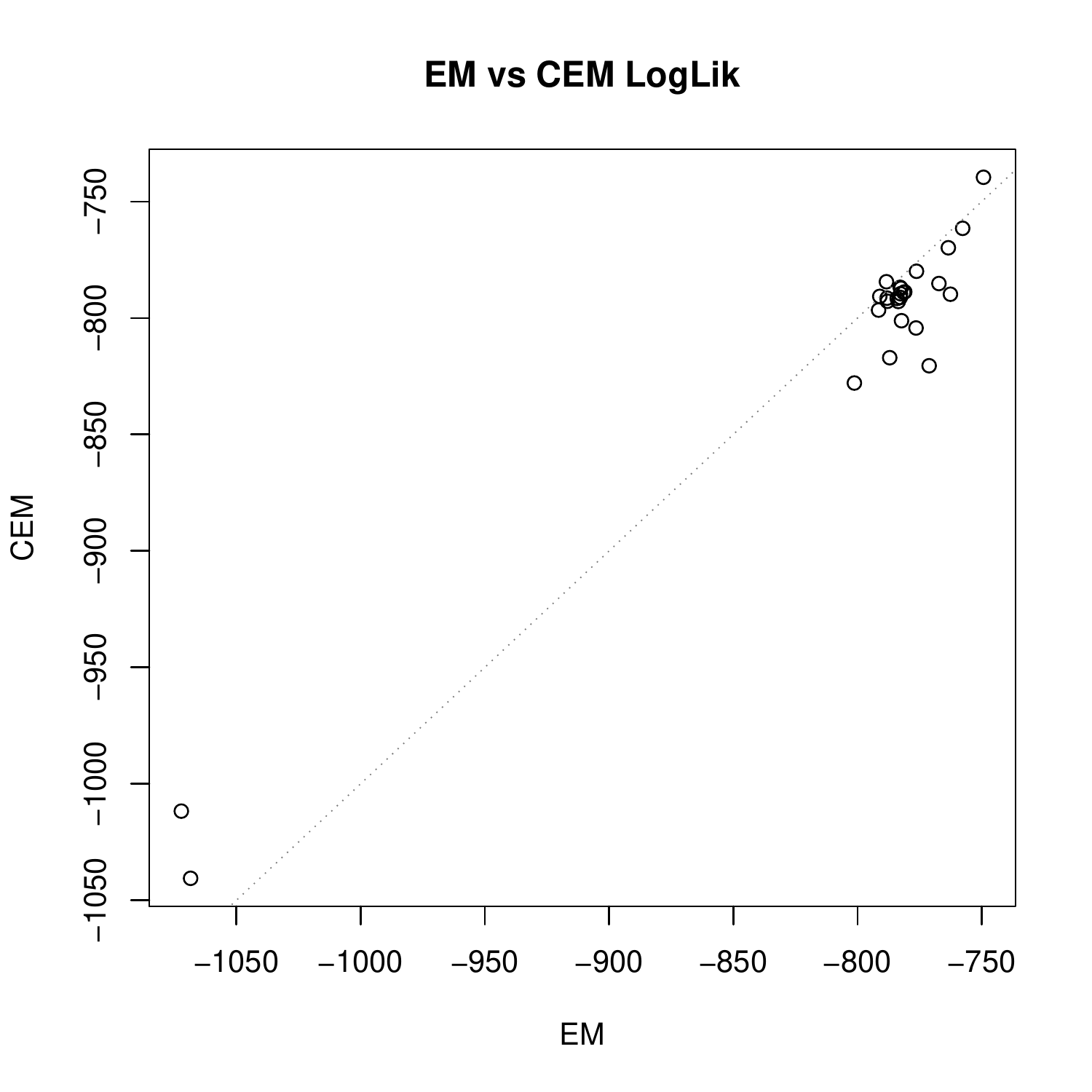}
\end{center}
\caption{Protein data set, cluster 2. First row: estimated means and $95\%$ ellipsoid confidence using optim, EM and CEM. Second row: comparison of the log-likelihood at the estimated values.}
\label{fig:protein:2}
\end{figure}

\begin{figure}
\begin{center}
\includegraphics[height=0.3\textwidth]{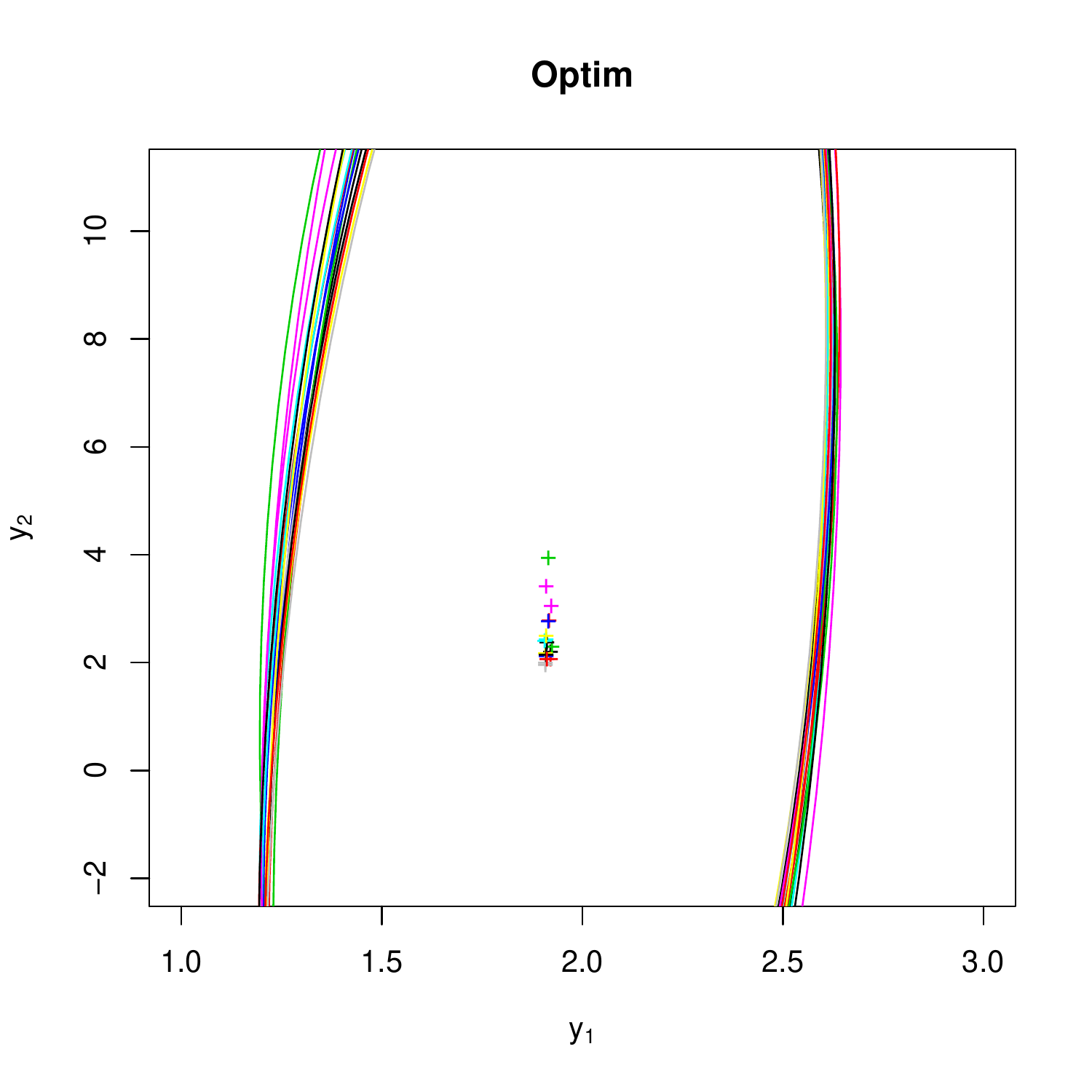}
\includegraphics[height=0.3\textwidth]{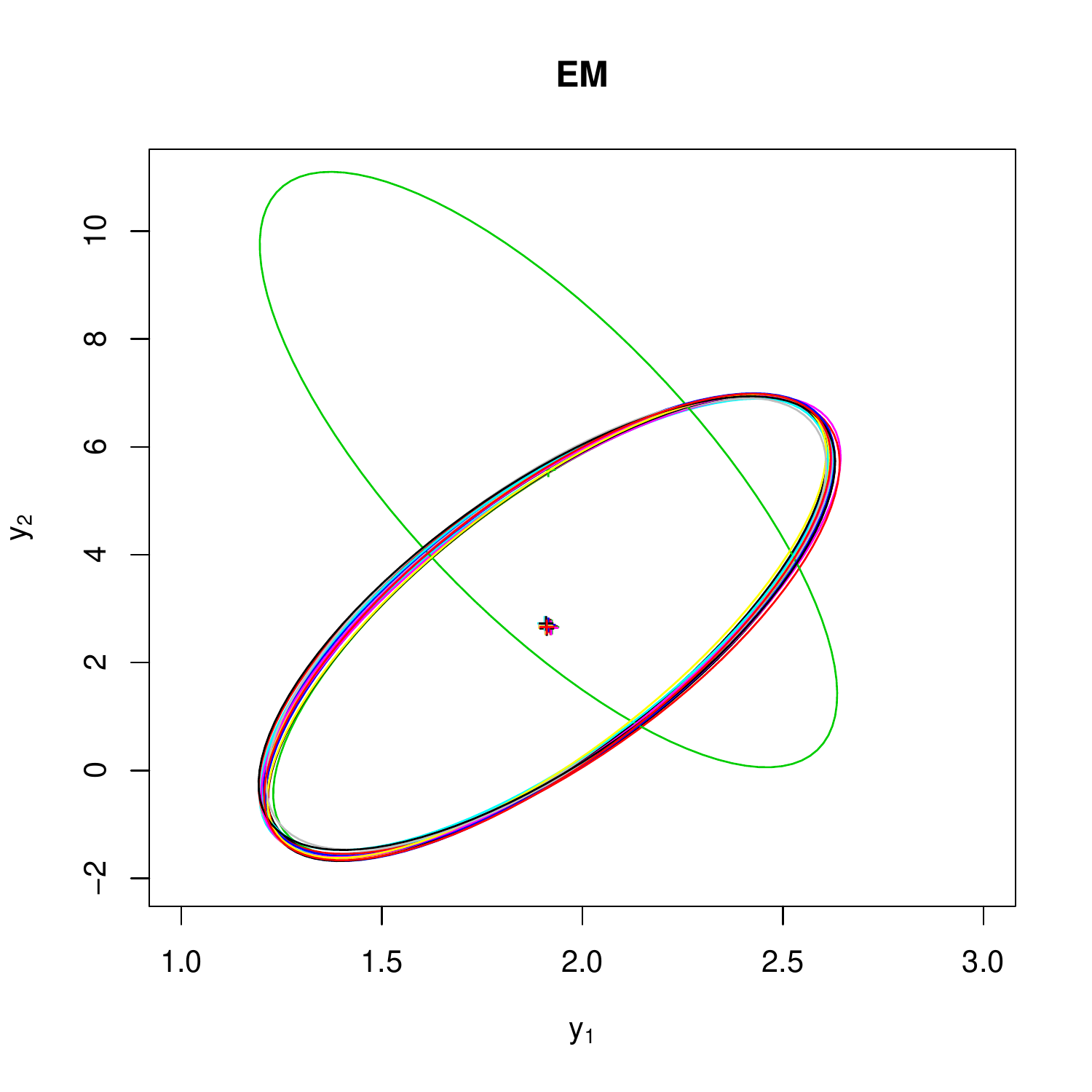}
\includegraphics[height=0.3\textwidth]{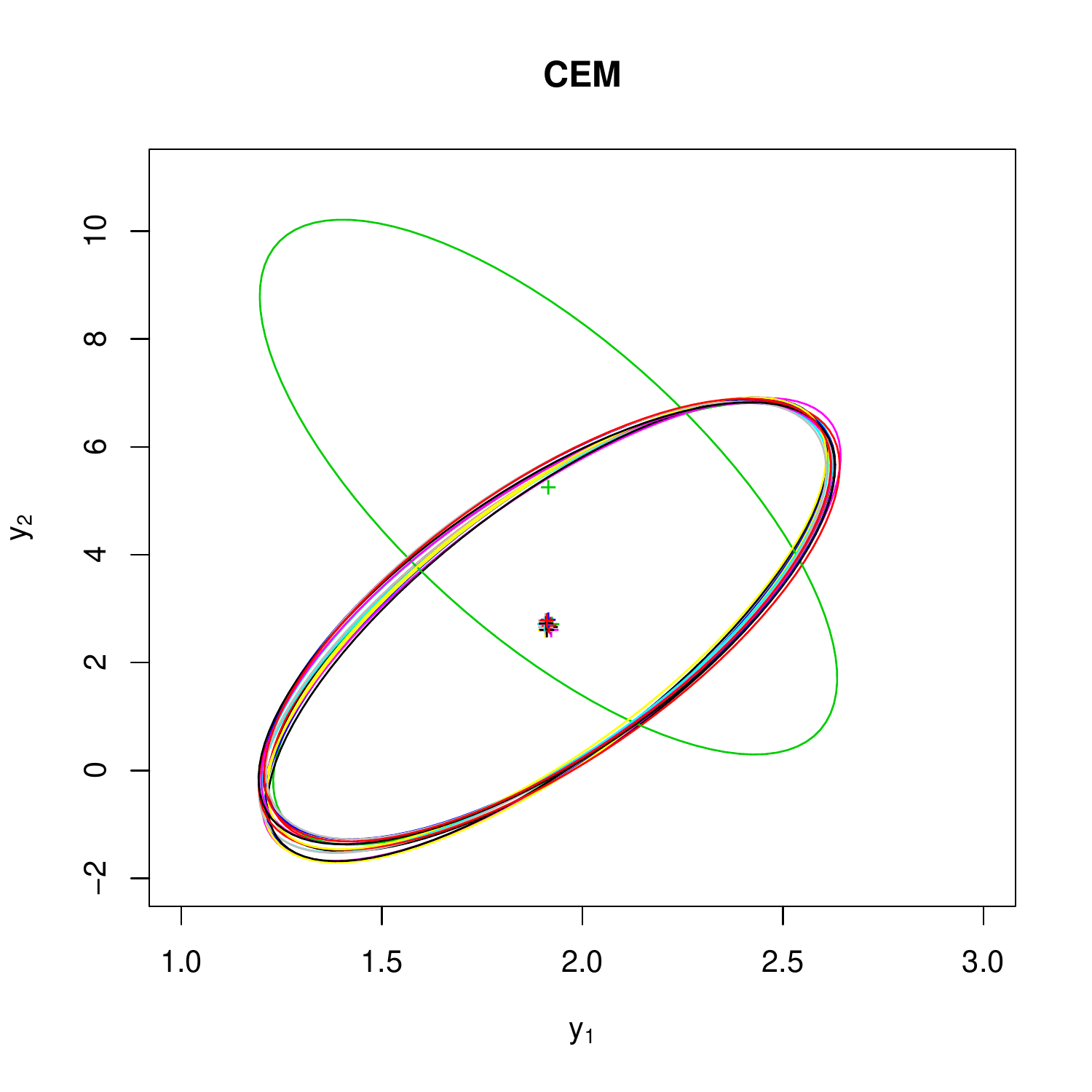} \\
\includegraphics[height=0.3\textwidth]{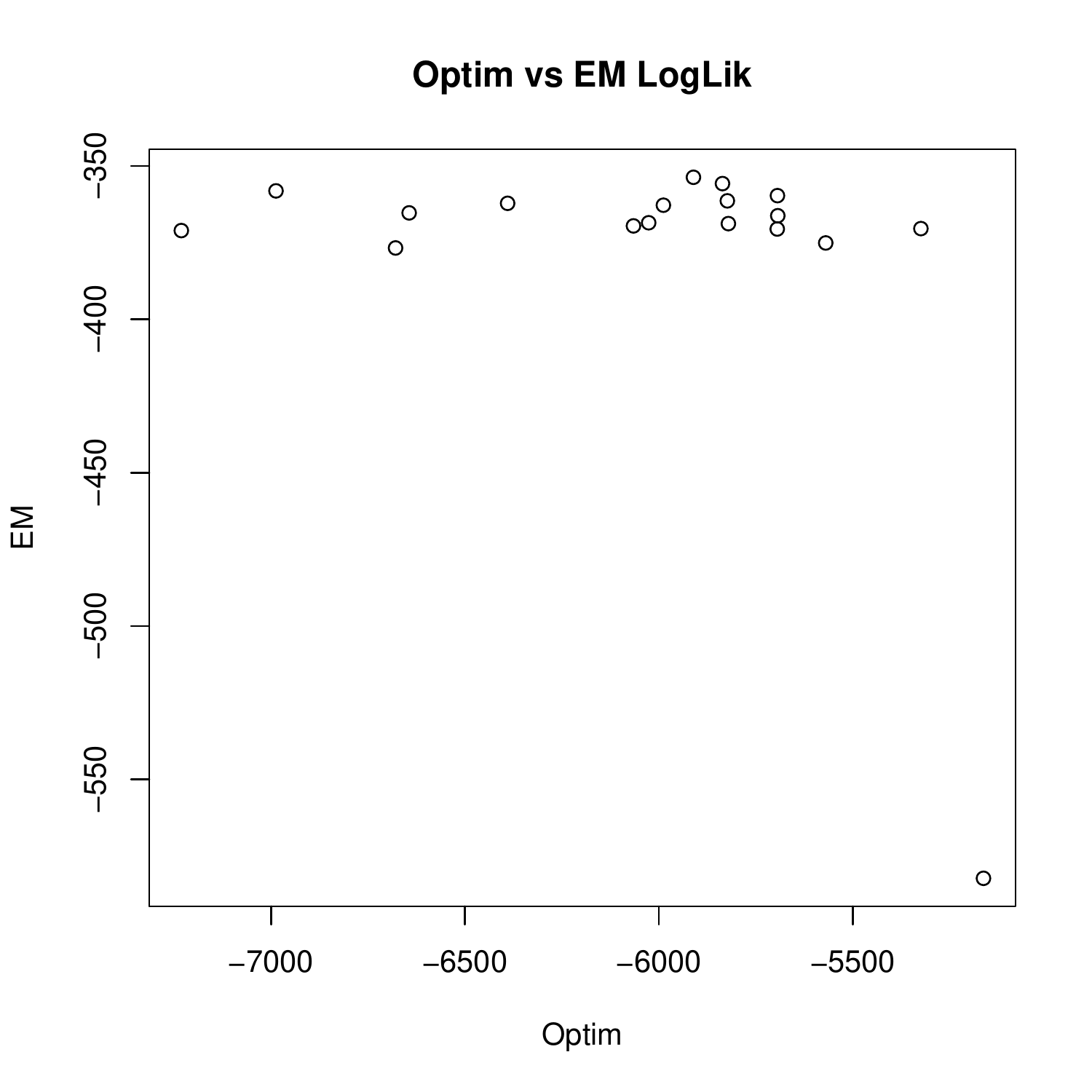}
\includegraphics[height=0.3\textwidth]{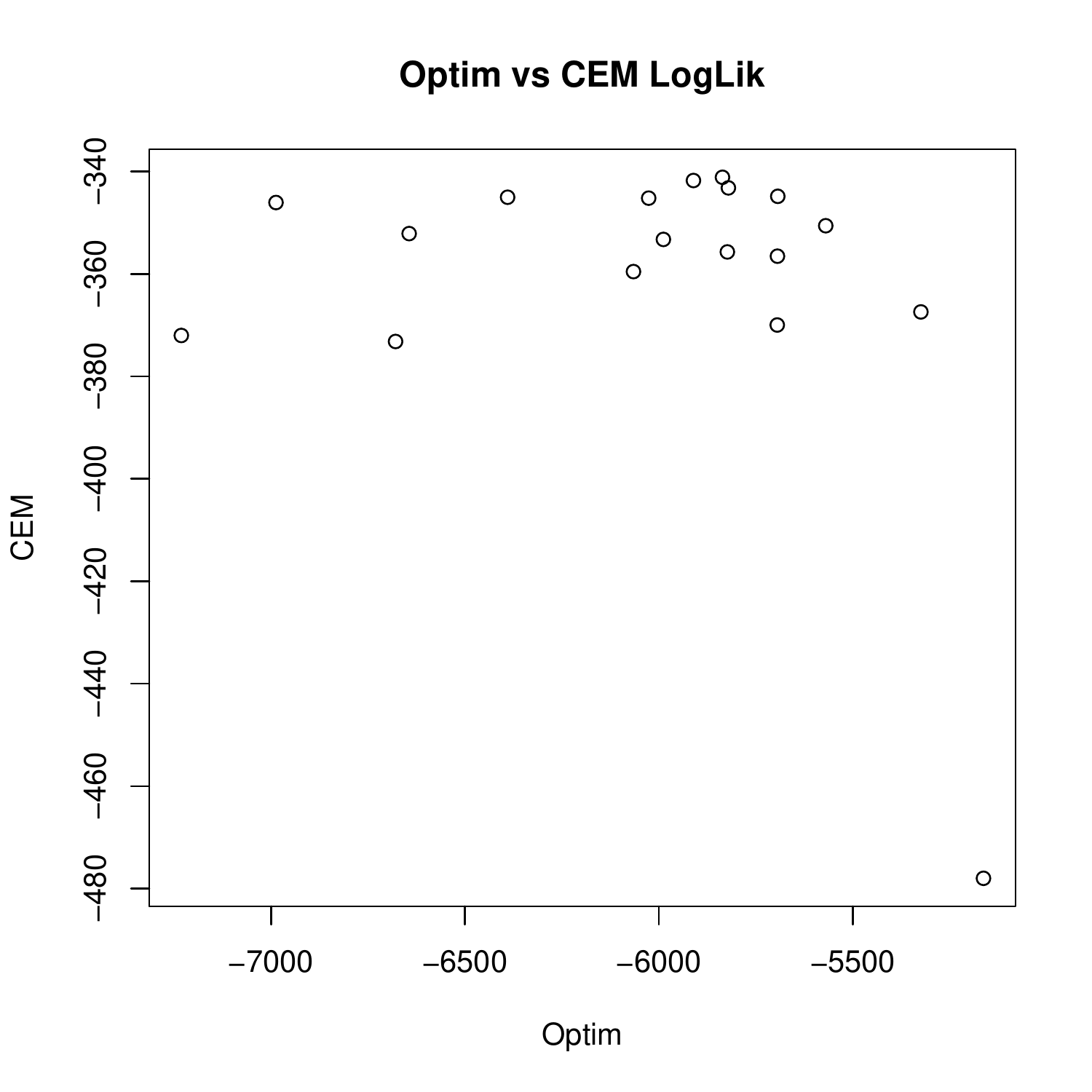}
\includegraphics[height=0.3\textwidth]{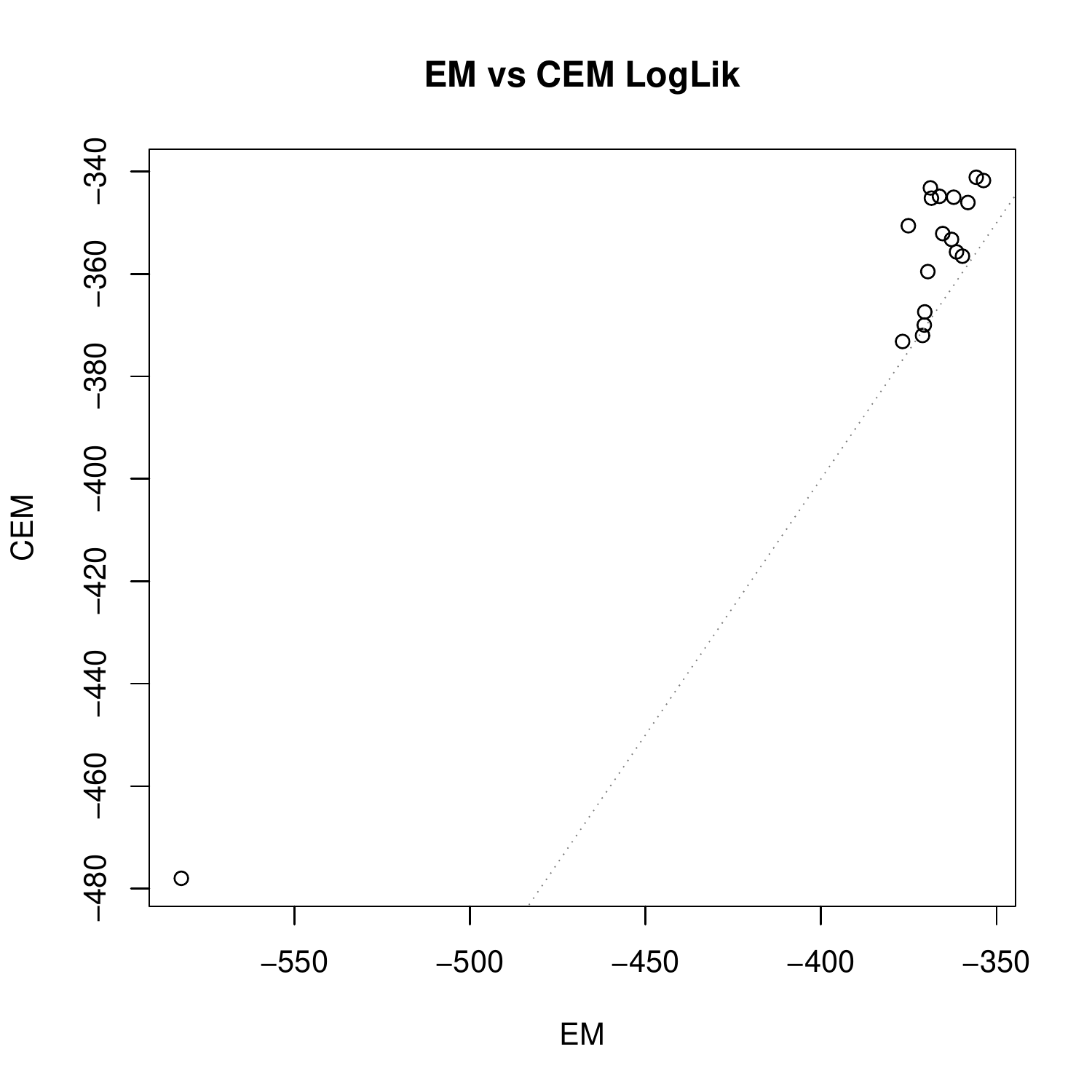}
\end{center}
\caption{Protein data set, cluster 3. First row: estimated means and $95\%$ ellipsoid confidence using optim, EM and CEM. Second row: comparison of the log-likelihood at the estimated values.}
\label{fig:protein:3}
\end{figure}

Here the protein domains are identified by their locations in the SCOP tree. Figures \ref{fig:protein:1}, \ref{fig:protein:2} and \ref{fig:protein:3} reports the results for the three clusters separately. For each Figure the first row shows the estimated means (+) and the $95\%$ confidence ellipsoid for the data based on the three considered procedures. The second row help on evaluating the performance of each method by comparing the log-likelihood at the estimates. From these plots we can see that, for almost all situations the EM and CEM report a larger value of the log-likelihood with respect to the direct optimization of the log-likelihood, this is particularly the case for Clusters 2 and 3. For those clusters, few data sets (one or two) show a dis-homogeneity with respect to the others.

\subsection{RNA data set: 7-torus}
\label{sec:rna}

In this example, we consider a data set on Ribonucleic acid (RNA). RNA is a polymeric molecule essential in various biological roles in coding, decoding, regulation, and expression of genes. RNA and DNA are nucleic acids, and, along with lipids, proteins and carbohydrates, constitute the four major macromolecules essential for all known forms of life. In RNA, each nucleic base corresponds to a backbone segment described by $6$ dihedral angles and one angle for the base, giving a total of $7$ angles. The distribution of these $7$ angles over large samples of RNA strands have been studied, among others, by \citet{Eltzner2018} using a Torus Principal Component Analysis. The original data set contains $8301$ observations, but based on a clustering procedure the data set was split in 23 clusters and all the observations with more than $50^{◦}$ in angular distance from their nearest neighbor removed. So, the final data set contains $7390$ observations grouped in 23 clusters. We apply the EM and CEM algorithms in each of these clusters to estimate the parameters and study the homogeneity of the groups. Due to moderate dimension on both sample size and number of variables, direct optimization of the log-likelihood become unfeasible and was not performed. In the Supplementary Material, in Tables SM-2 and SM-3 we reported estimated mean angles for each of the 23 clusters based on EM and CEM algorithm respectively. Also, Figures  SM-1 -- SM-6  represents the estimated correlation structures for all the $23$ clusters.

As seen in both Tables SM-2 and SM-3, the mean estimation in each cluster is close to each other. Also,  Figures  SM-1 -- SM-6  depict each correlation by an ellipse whose shape tends towards a line with slope $1$ for correlations near $1$, to a circle for correlations near zero, and to a line with negative slope $−1$ for negative correlations near $−1$. In addition, a colour indicates strong negative (red) to strong positive (blue) correlations. An inspection of these plots shows a nice agreement of the two (EM and CEM) algorithms.

\section{Monte Carlo experiments}
\label{sec:simulations}

To compare the performance of the proposed methods we consider two Monte Carlo experiments, the first for the univariate case; the second for the multivariate case. In the univariate case the experiment has the following factors: sample size $n=10, 50, 100, 500$, $\mu_0 = 0$, $\sigma_0=(\pi/8, \pi/4, \pi/2, \pi, 3/2\pi, 2\pi)$, and number of Monte Carlo replications $500$. We compare the following methods, EM, CEM algorithms, direct maximization of the log-likelihood perfomed using the R function \texttt{optim} with default values. As initial values for all these methods we use the circular mean and $-2\log(\hat{\rho})$ respectively for $\mu$ and $\sigma$, where $\hat{\rho}$ is the sample mean resultant length. We also consider the algorithm implemented in function \texttt{mle.wrappednormal} available in the R package \texttt{circular} \citep{AgostinelliLund2017} which is based on an iterative re-weighting algorithm that computes the maximum likelihood estimate. 

In the multivariate case the experiment has the following factors: number of variables $p=2, 5, 10$, and sample size depending on $p$ in the range $n=50, 100, 500$, $\vect{\mu}_0 = \vect{0}$ and number of Monte Carlo replications $500$.

To account for the lack of affine equivariance of the wrapped model we consider different covariance structures $\Sigma_0$ as in \citet{AgostinelliEtAll2015}. For each sample in our simulation we create a different random correlation matrix with condition number fixed at $\text{CN} = 20$. We use the following procedure to obtain random correlations with a fixed condition number $\text{CN}$:

\begin{enumerate}
\item For a fixed condition number $\text{CN}$, we first obtain a diagonal matrix $\Lambda = \diag(\lambda_1, \ldots, \lambda_p),$  ($\lambda_1 < \lambda_2 < \ldots < \lambda_p$) with smallest eigenvalue $\lambda_1 = 1$ and largest eigenvalue $\lambda_p = \text{CN}$. The remaining eigenvalues $\lambda_2, \ldots, \lambda_{p-1}$ are $(p-2)$ sorted independent random variables with a uniform distribution in the interval $\left(1, \text{CN} \right)$.

\item We first generate a random $p \times p$ matrix $Y$, in which its elements are independent standard normal random variables. Then we form the symmetric matrix $Y^\top Y = U V U^\top$ to obtain a random orthogonal matrix $U$.

\item Using the results of 1 and 2 above, we construct the random covariance matrix by $\Sigma =  U \Lambda U^\top$. Notice that the condition number of $\Sigma$ is equals to the desired $\text{CN}$.

\item Convert the covariance matrix $\Sigma$ into the correlation matrix $R$ as follows:
\begin{equation*}
R = D^{-1/2} \Sigma D^{-1/2}
\end{equation*}
where
\begin{equation*}
D = \diag(d_1, \ldots, d_p) \, 
\end{equation*}
and $(d_1,\ldots,d_p)$ are the variances of $\Sigma$. 
\item After the conversion to correlation matrix in step 4 above, the condition number of $R$ is no longer necessarily equal to $\text{CN}$. To remedy this problem, we consider the eigenvalue diagonalization of $R$ 
\begin{equation}\label{eq:MC-Correlation}
R = U \Lambda U^\top \ , 
\end{equation}
where
\begin{equation*}
\Lambda = \diag(\lambda_1^{(R)}, \ldots, \lambda_p^{(R)}), \qquad \lambda_1^{(R)} < \lambda_2^{(R)} \ldots < \lambda_p^{(R)}
\end{equation*}
is the diagonal matrix formed using the eigenvalues of $R$. We now re-establish the desired condition number $CN$ by redefining 
\begin{equation*}
\lambda_p^{(R)} = \text{CN} \times \lambda_1^{(R)}
\end{equation*}
and using the modified eigenvalues in (\ref{eq:MC-Correlation}).

\item Repeat 4 and 5 until the condition number of $R$ is within a tolerance level (or until we reach some maximum iterations). In our Monte Carlo study convergence was reached after a few iteration in all the cases.
\end{enumerate}
Once a desidered correlation matrix is obtained, covariance matrices are considered so that variances in the main diagonal are the square of $\sigma_0$, chosen  among the values $(\pi/8, \pi/4, \pi/2, \pi, 3/2\pi, 2\pi)$ as in the univariate case.

We compare the following methods, EM, CEM algorithms, direct maximization of the log-likelihood \eqref{equ:loglik} perfomed using the R function \texttt{optim} with the default setting. As initial values for all these methods we use the approach reported in Section \ref{sec:initial} which proved to be fast and effective. We also start the algorithms from the true values $\vect{\mu}_0$ and $\Sigma_0$ in order to better understand the effect of the initial values in the performance of the methods. A ``T'' is added at the end of the labels for these cases. For evaluation of the log-likelihood, the covariance matrix $\Sigma$ is parametrized using the Log-Cholesky parameterization \citep{PinheiroBates1996} as described in Section \ref{sec:estimation} which allows for uncostrained optimization while ensuring positive definite estimate of $\Sigma$.

In all cases the performance is evaluated using three measures: (i) the Wilks' test statistics based on log-likelihood, i.e.
\begin{equation*}
\Lambda(\hat{\vect{\mu}}, \hat{\Sigma}) = -2 (\ell(\vect{\mu}_0, \Sigma_0) - \ell(\hat{\vect{\mu}}, \hat{\Sigma})) \ ,
\end{equation*}
where $\vect{\mu}_0$ and $\Sigma_0$ are the true values;
(ii) the angle separation:
\begin{equation*}
\AS(\hat{\vect{\mu}}) = \sum_{i=1}^p (1 - \cos(\hat{\mu}_i - \mu_{i0})) \ , 
\end{equation*}
with a range in $[0, 2\ p]$;
(iii) the performance of a given scatter estimator $\hat{\Sigma}$ between two Gaussian distribution with the same mean and covariances $\hat{\Sigma}$ and $\Sigma_0$ can be evaluated using:
\begin{equation*}
\Delta(\hat{\Sigma}) = \trace(\hat{\Sigma} \Sigma_0^{-1}) - \log(| \hat{\Sigma} \Sigma_0^{-1} |) - p \ .
\end{equation*}
This divergence also appears in the likelihood ratio test statistics for testing the null hypothesis that a multivariate normal distribution has covariance matrix $\Sigma = \Sigma_0$.

\begin{figure}
\begin{center}
\includegraphics[height=0.3\textheight]{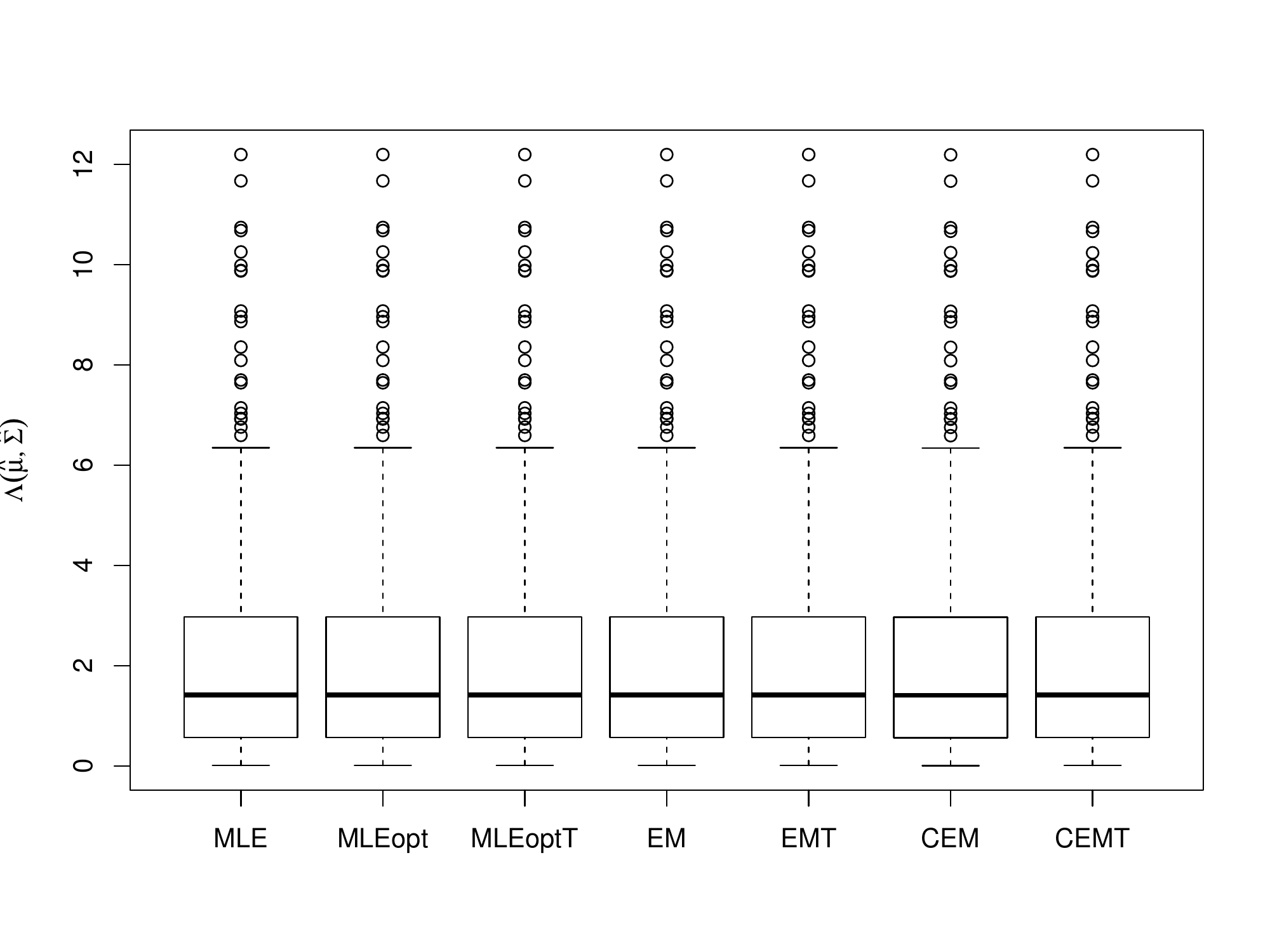} \\
\includegraphics[height=0.3\textheight]{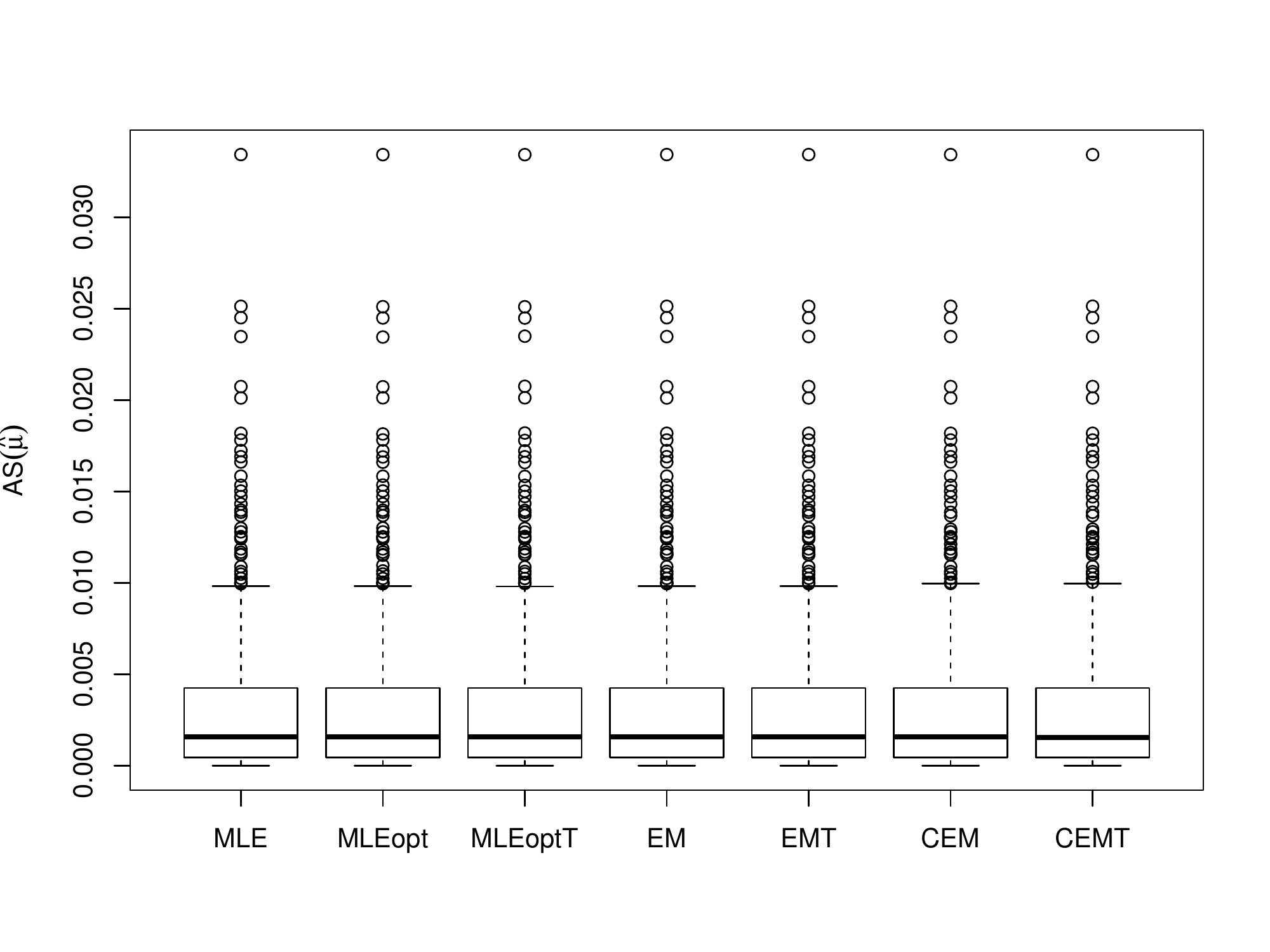} \\
\includegraphics[height=0.3\textheight]{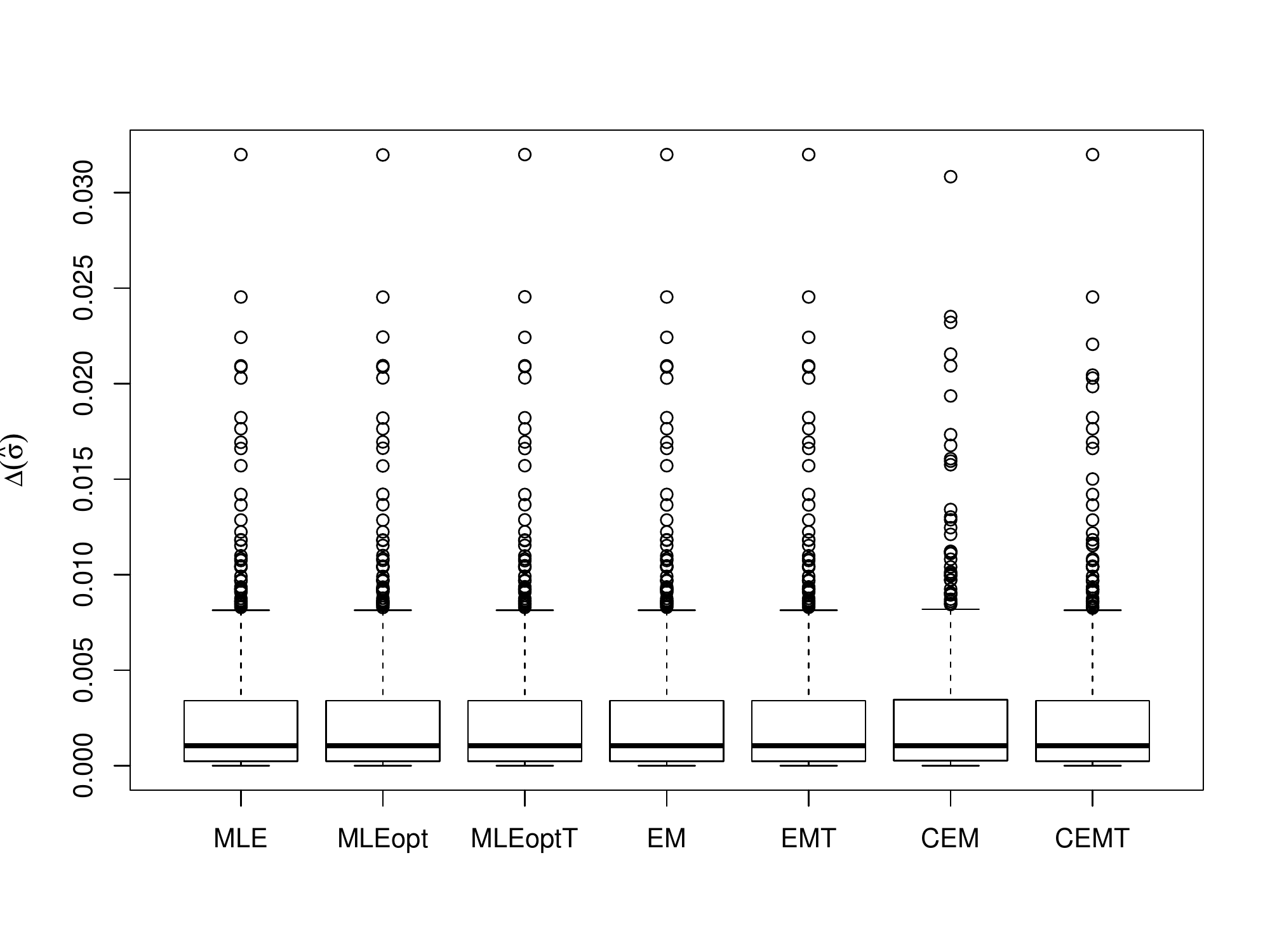} \\
\end{center}
\caption{Performance of the estimators in the univariate case $p=1$, sample size $n=100$, $\sigma=\pi/4$.}
\label{fig:1:100:2}
\end{figure}

\begin{figure}
\begin{center}
\includegraphics[height=0.3\textheight]{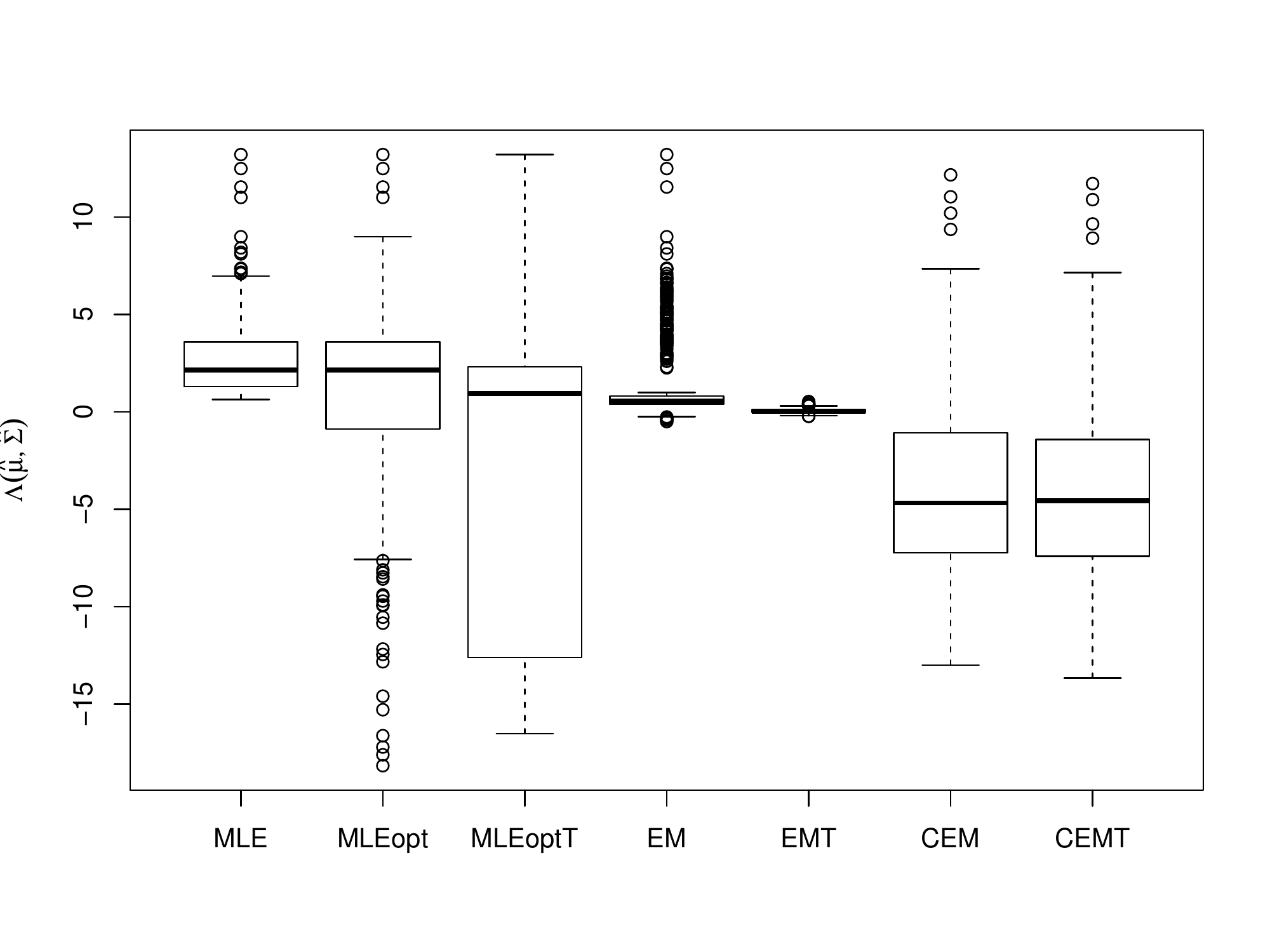} \\
\includegraphics[height=0.3\textheight]{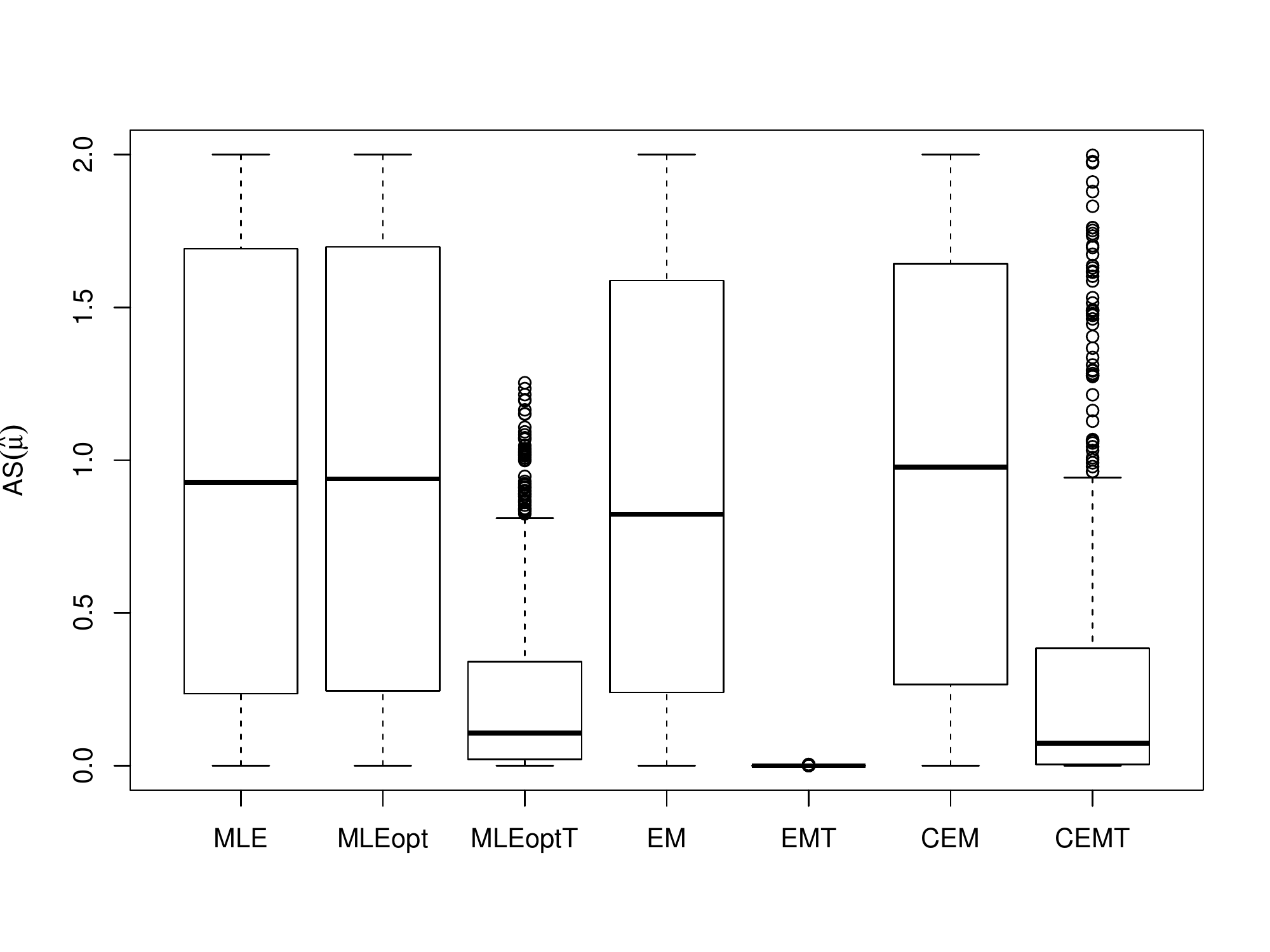} \\
\includegraphics[height=0.3\textheight]{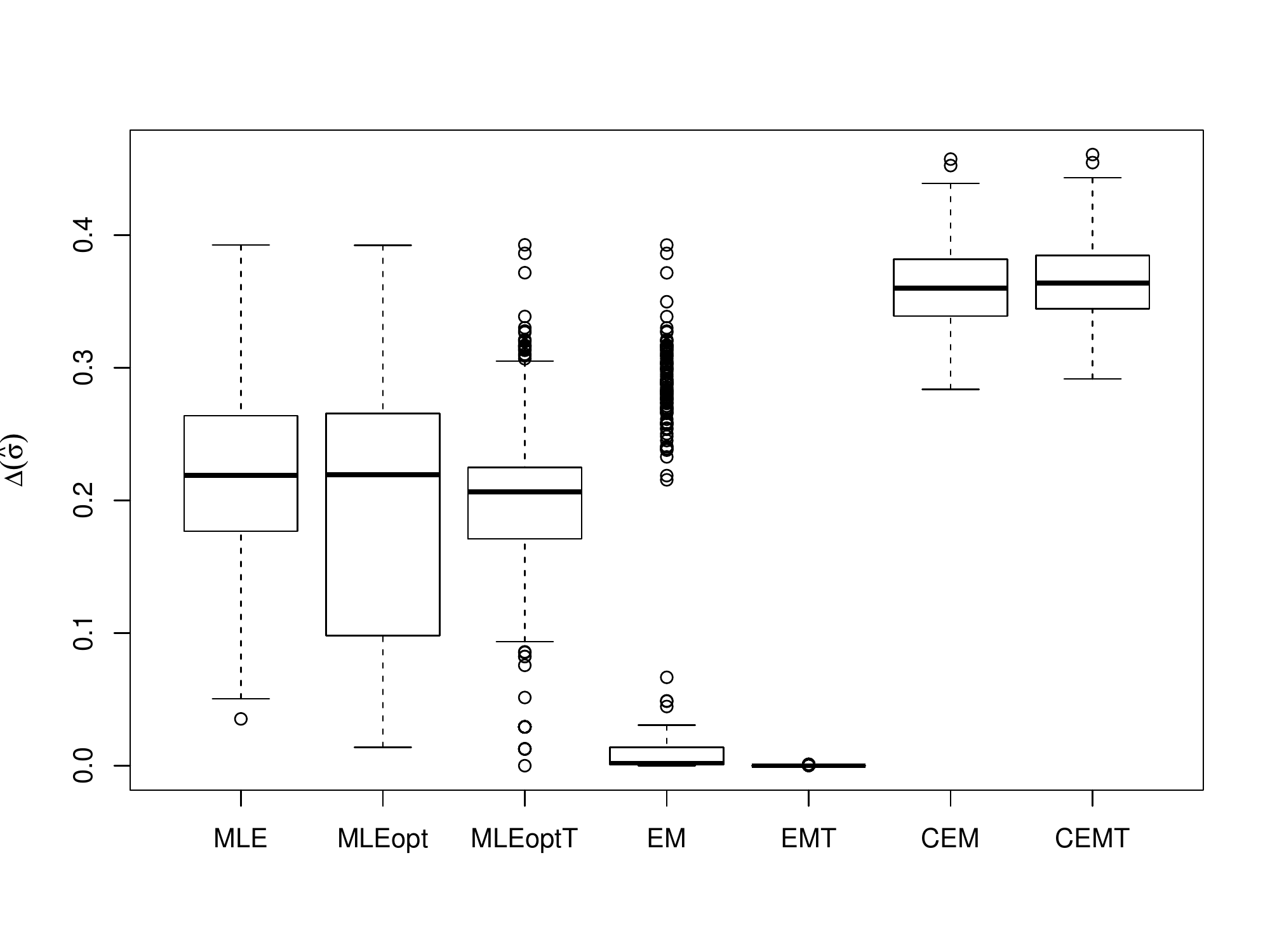} \\
\end{center}
\caption{Performance of the estimators in the univariate case $p=1$, sample size $n=100$, $\sigma=3 \pi/2$.}
\label{fig:1:100:5}
\end{figure}

\begin{figure}
\begin{center}
\includegraphics[height=0.3\textheight]{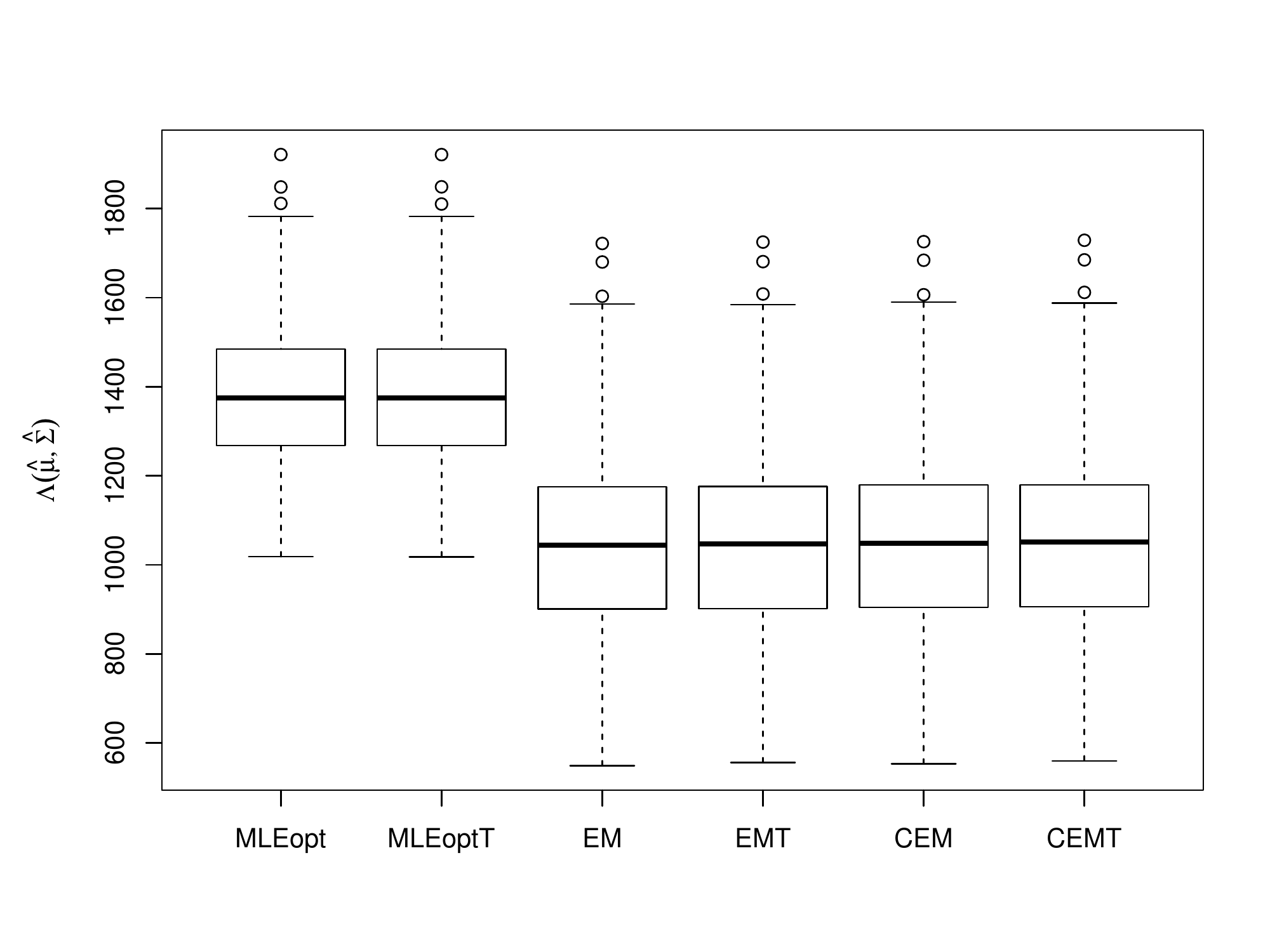} \\
\includegraphics[height=0.3\textheight]{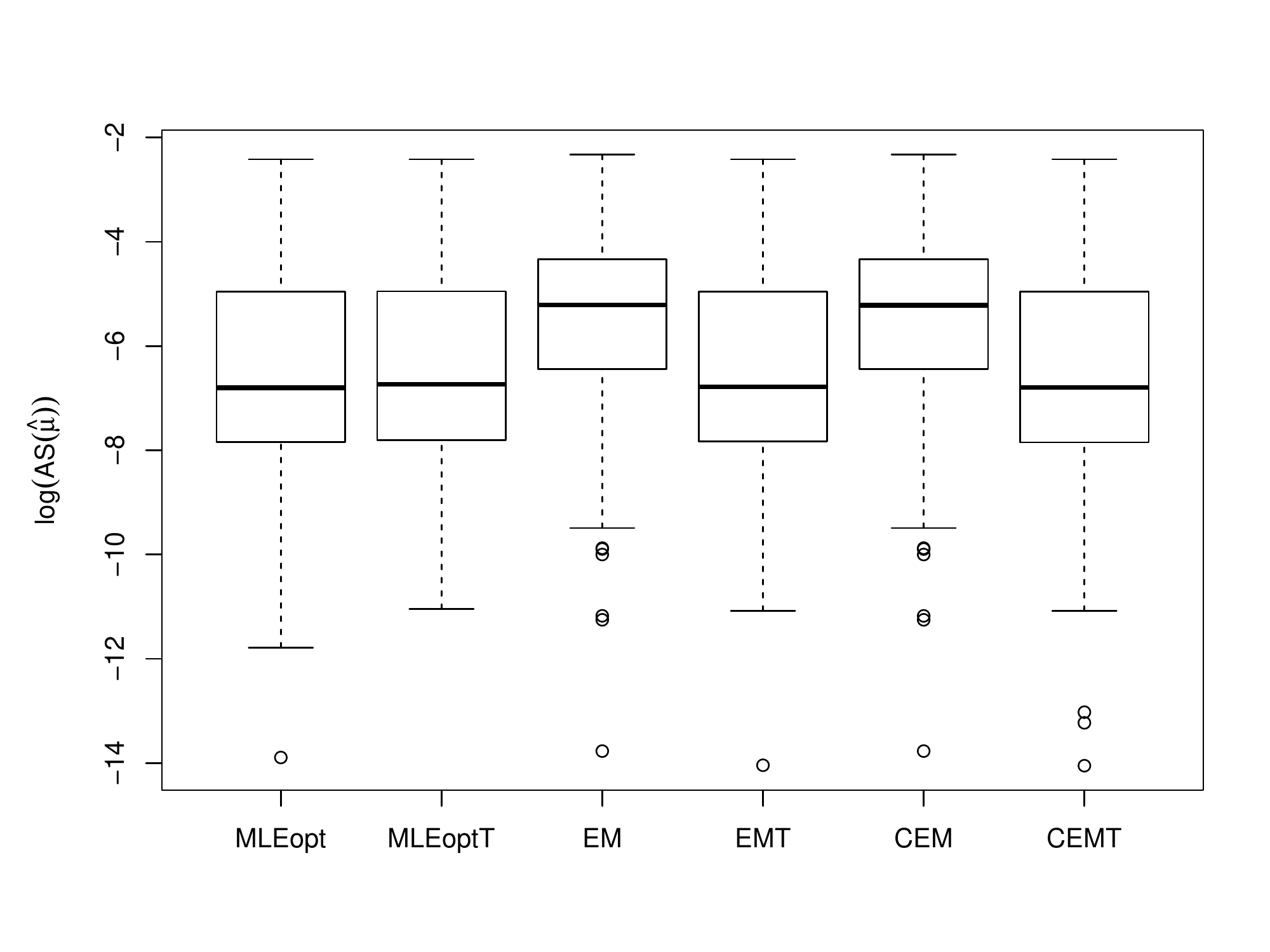} \\
\includegraphics[height=0.3\textheight]{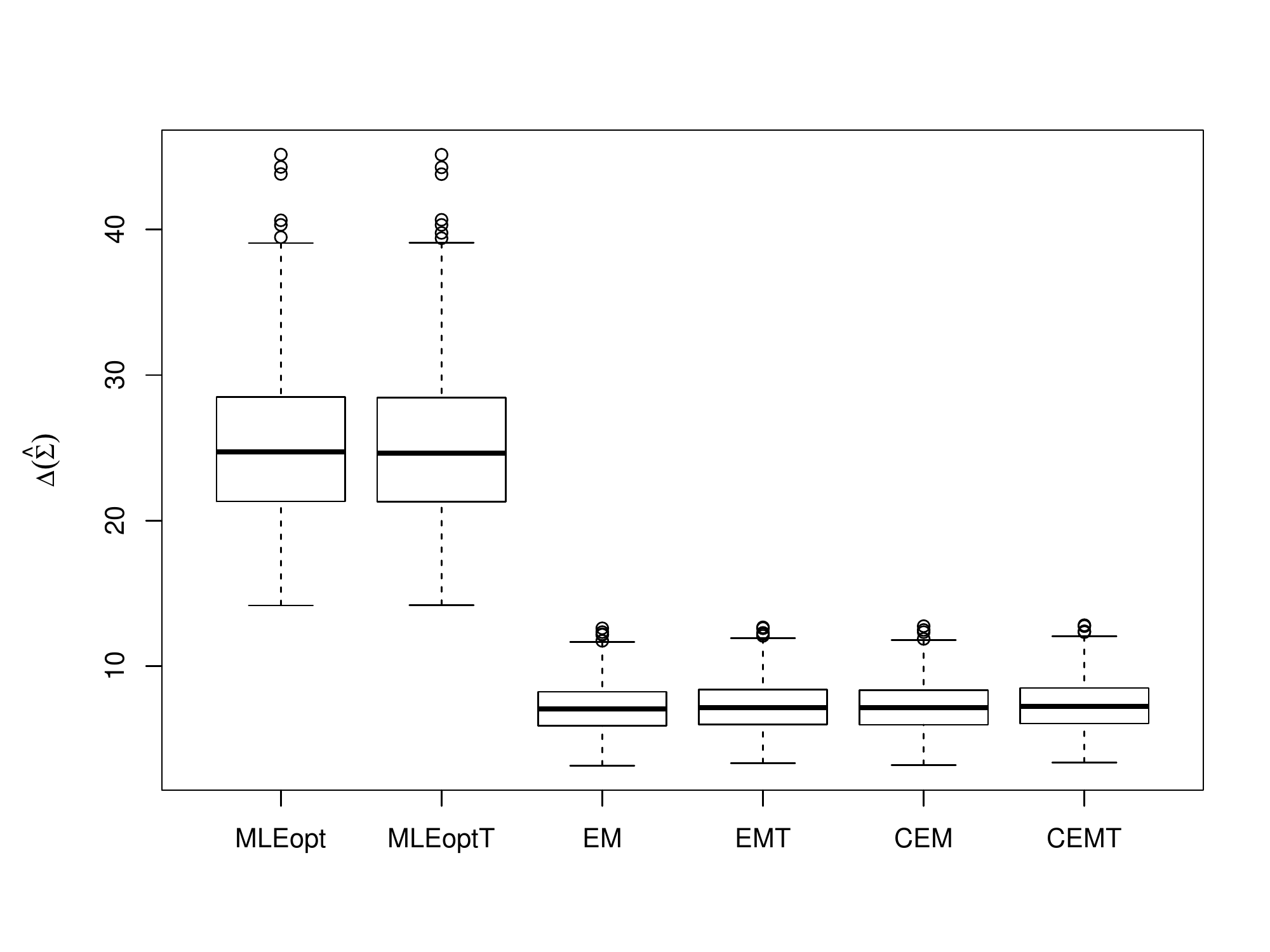} \\
\end{center}
\caption{Performance of the estimators in the bivariate case $p=2$, sample size $n=100$, $\sigma=\pi/4$.}
\label{fig:2:100:2}
\end{figure}

\begin{figure}
\begin{center}
\includegraphics[height=0.3\textheight]{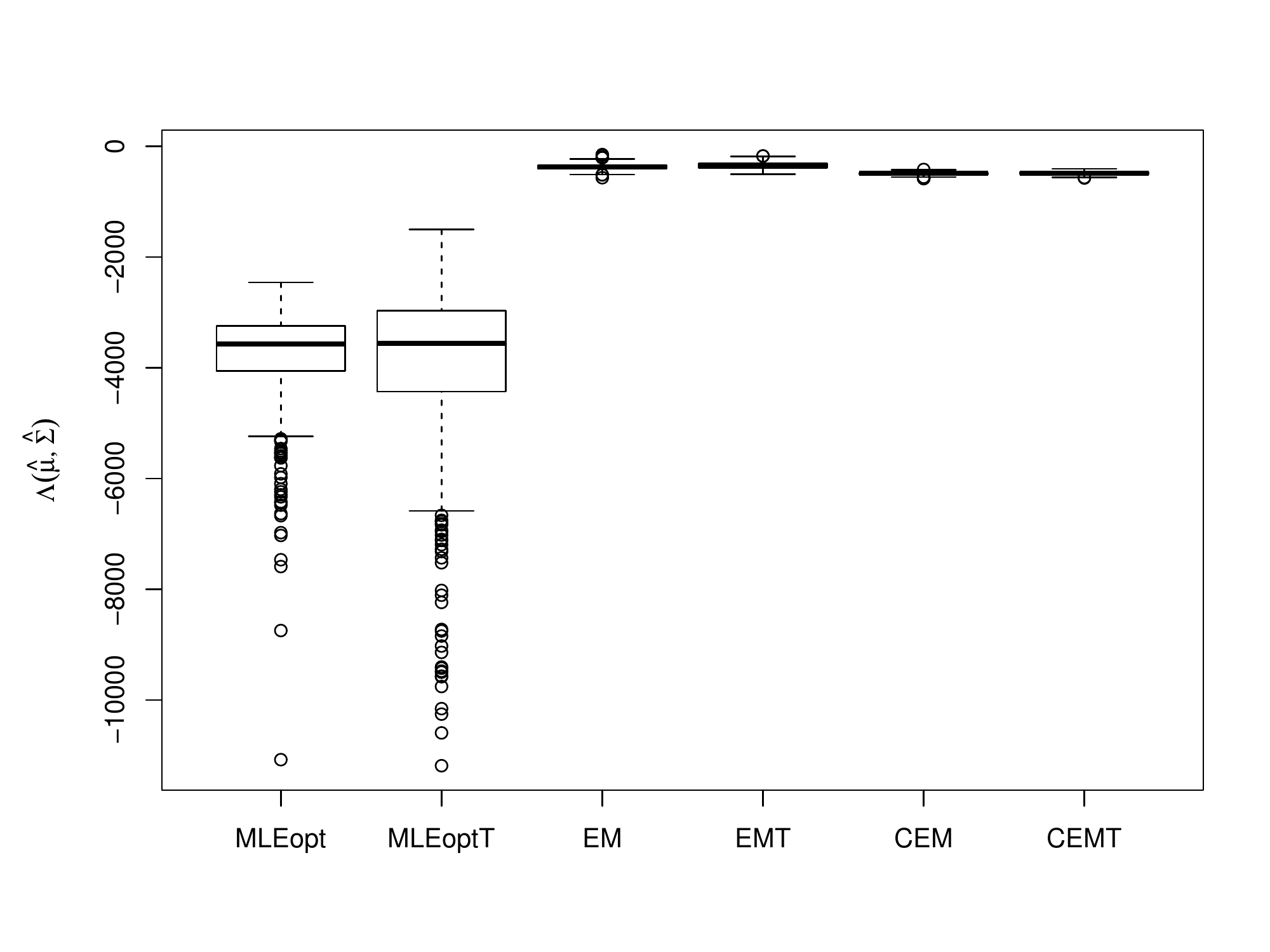} \\
\includegraphics[height=0.3\textheight]{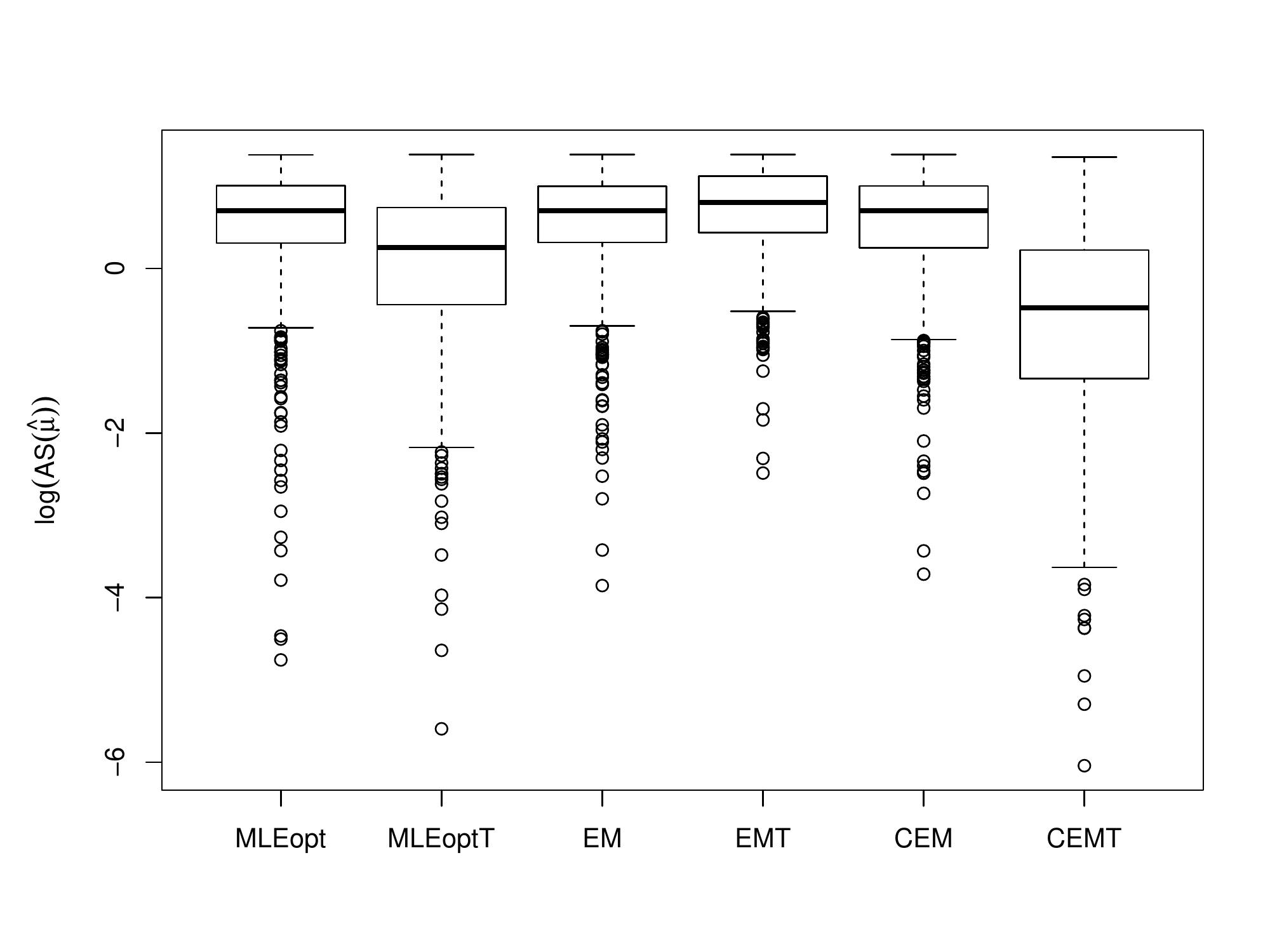} \\
\includegraphics[height=0.3\textheight]{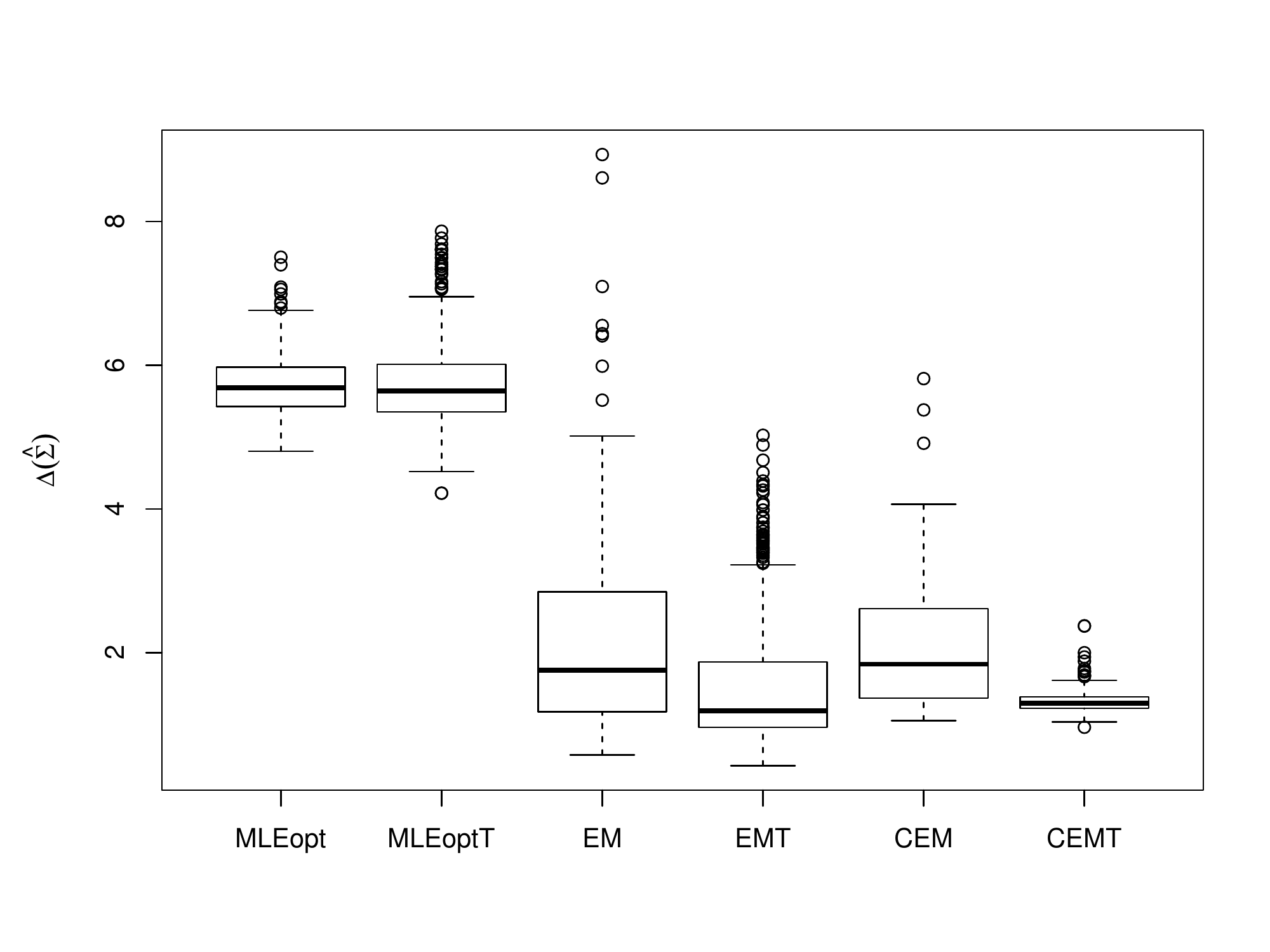} \\
\end{center}
\caption{Performance of the estimators in the bivariate case $p=2$, sample size $n=100$, $\sigma=3 \pi/2$.}
\label{fig:2:100:5}
\end{figure}

\begin{figure}
\begin{center}
\includegraphics[height=0.3\textheight]{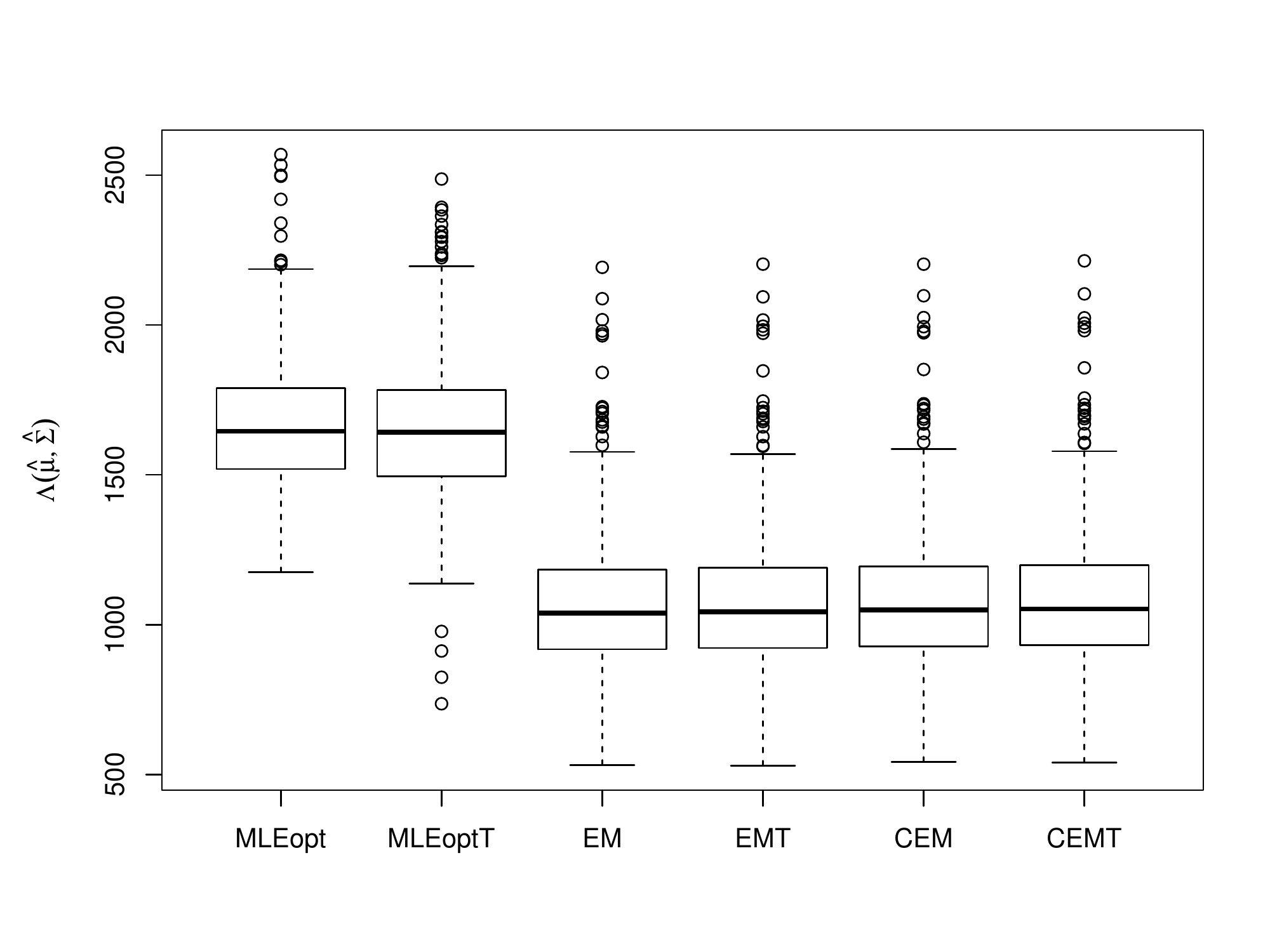} \\
\includegraphics[height=0.3\textheight]{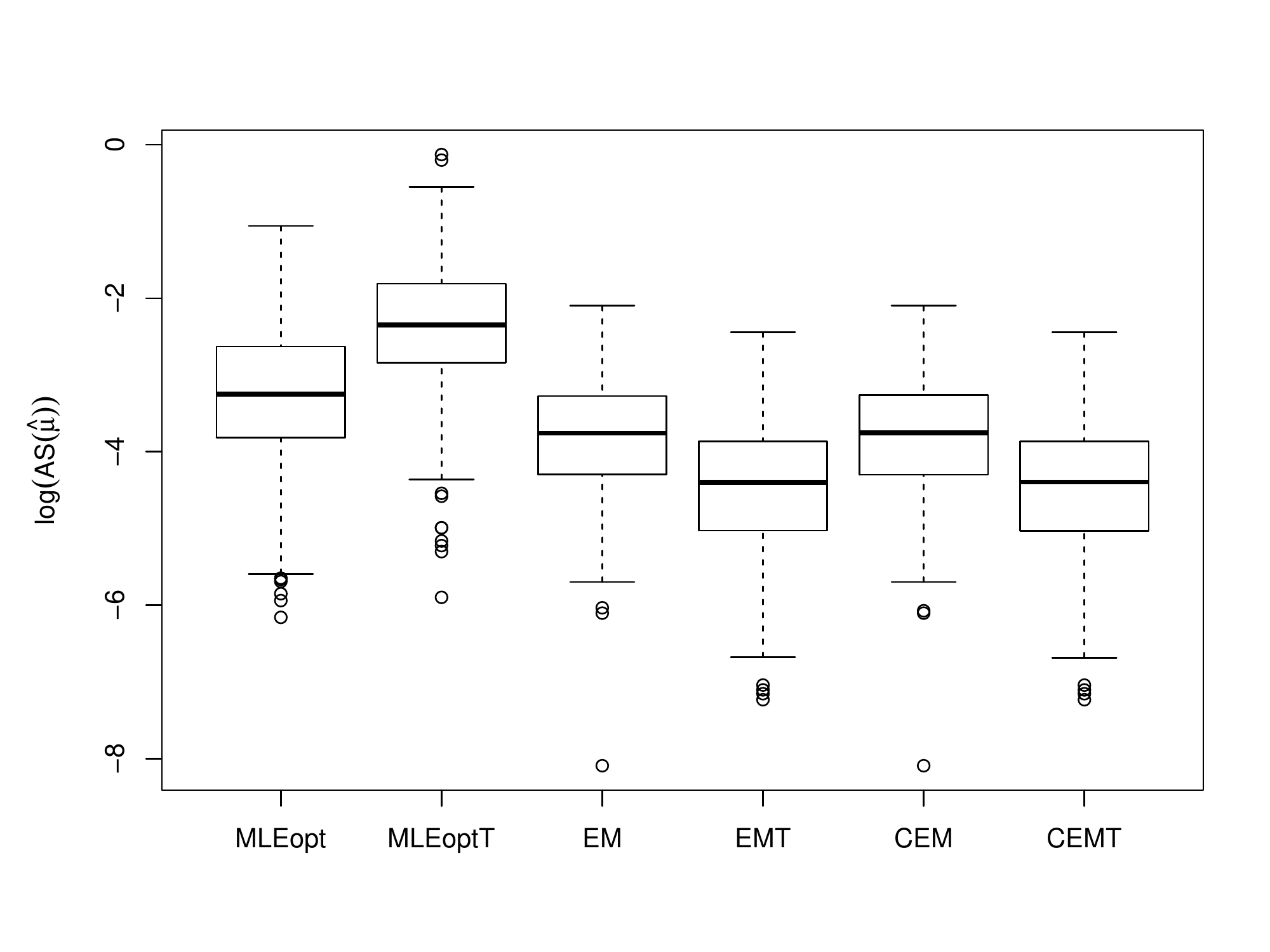} \\
\includegraphics[height=0.3\textheight]{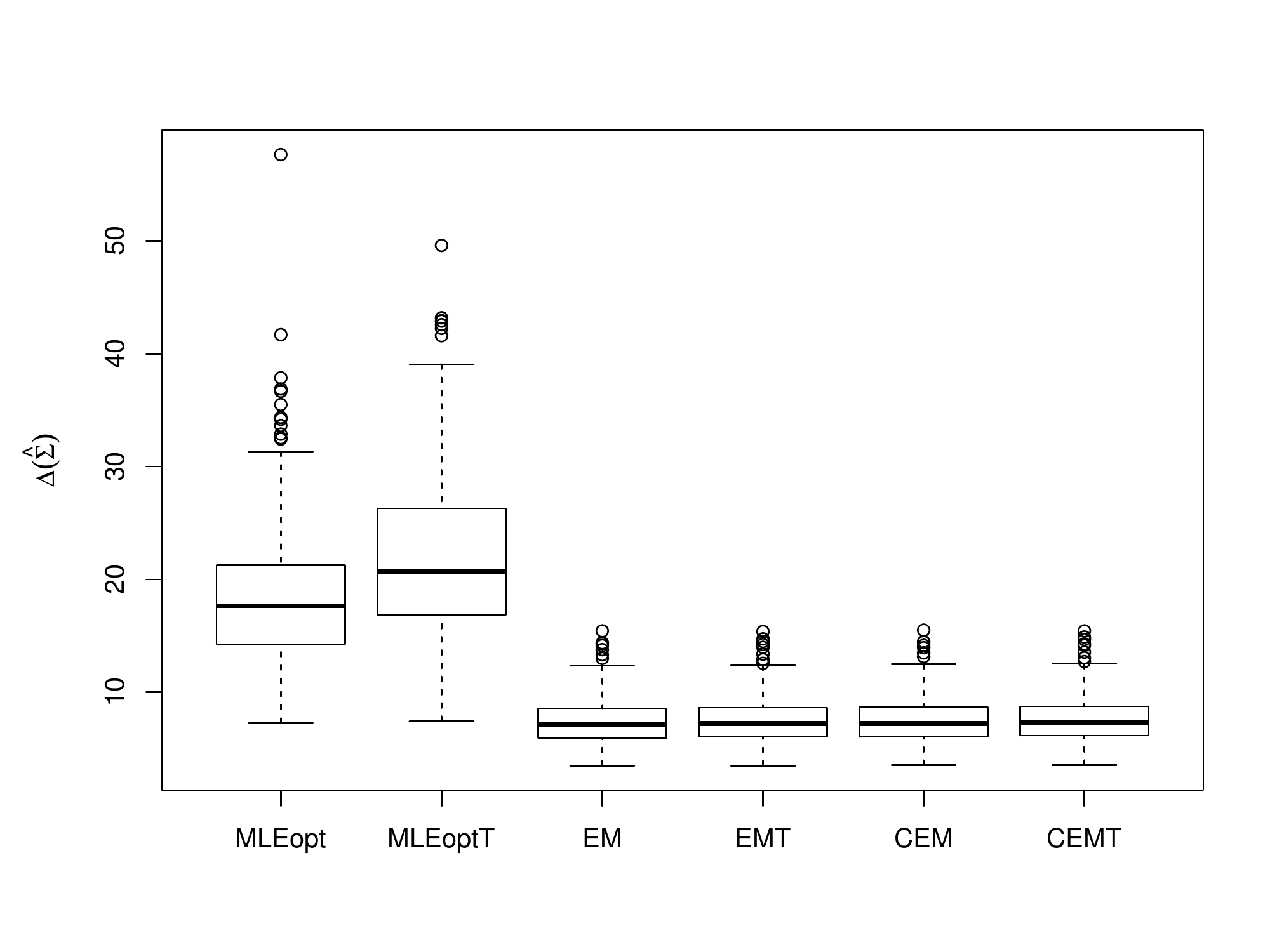} \\
\end{center}
\caption{Performance of the estimators in the case $p=5$, sample size $n=100$, $\sigma=\pi/4$.}
\label{fig:5:100:2}
\end{figure}

\begin{figure}
\begin{center}
\includegraphics[height=0.3\textheight]{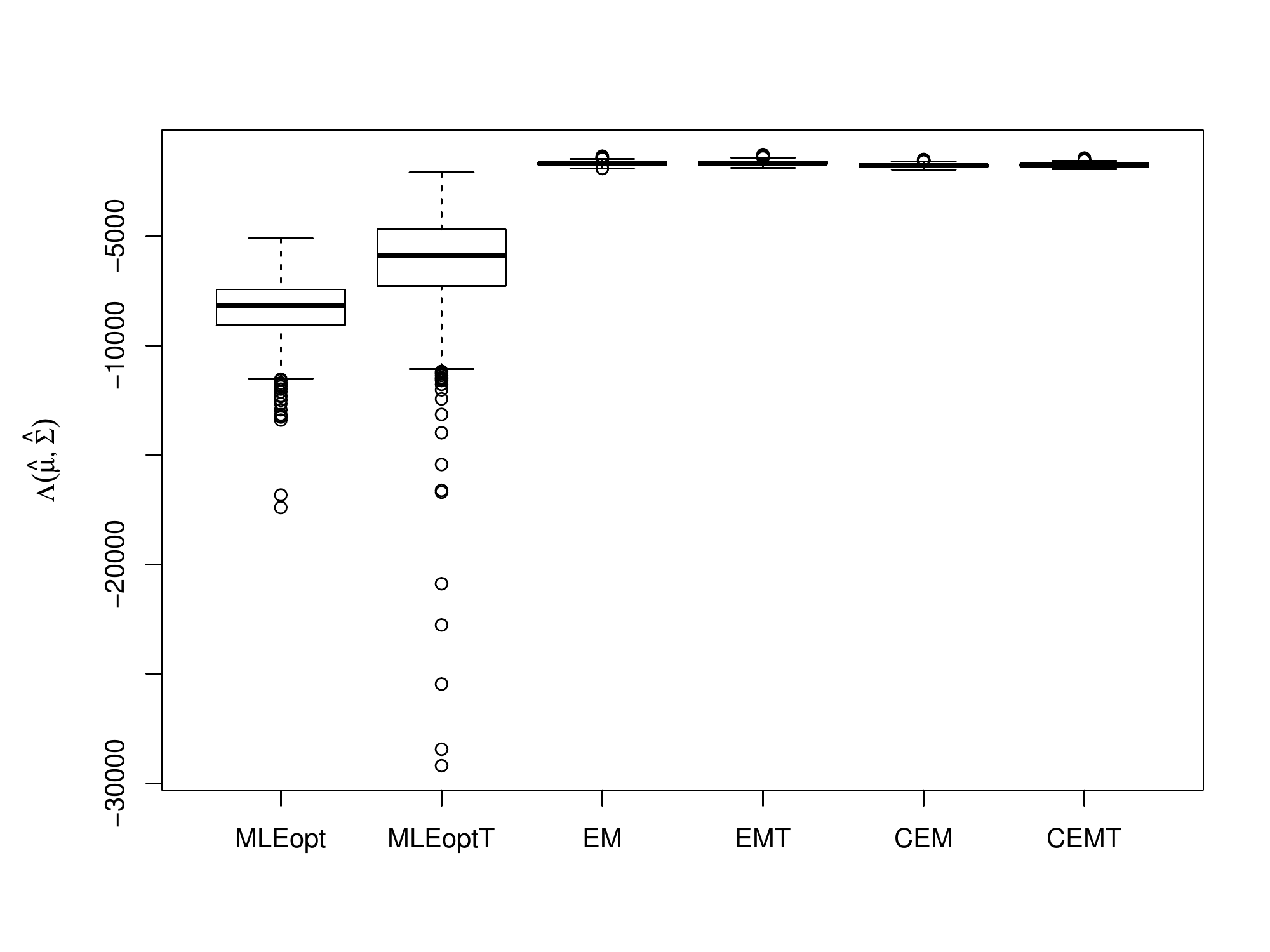} \\
\includegraphics[height=0.3\textheight]{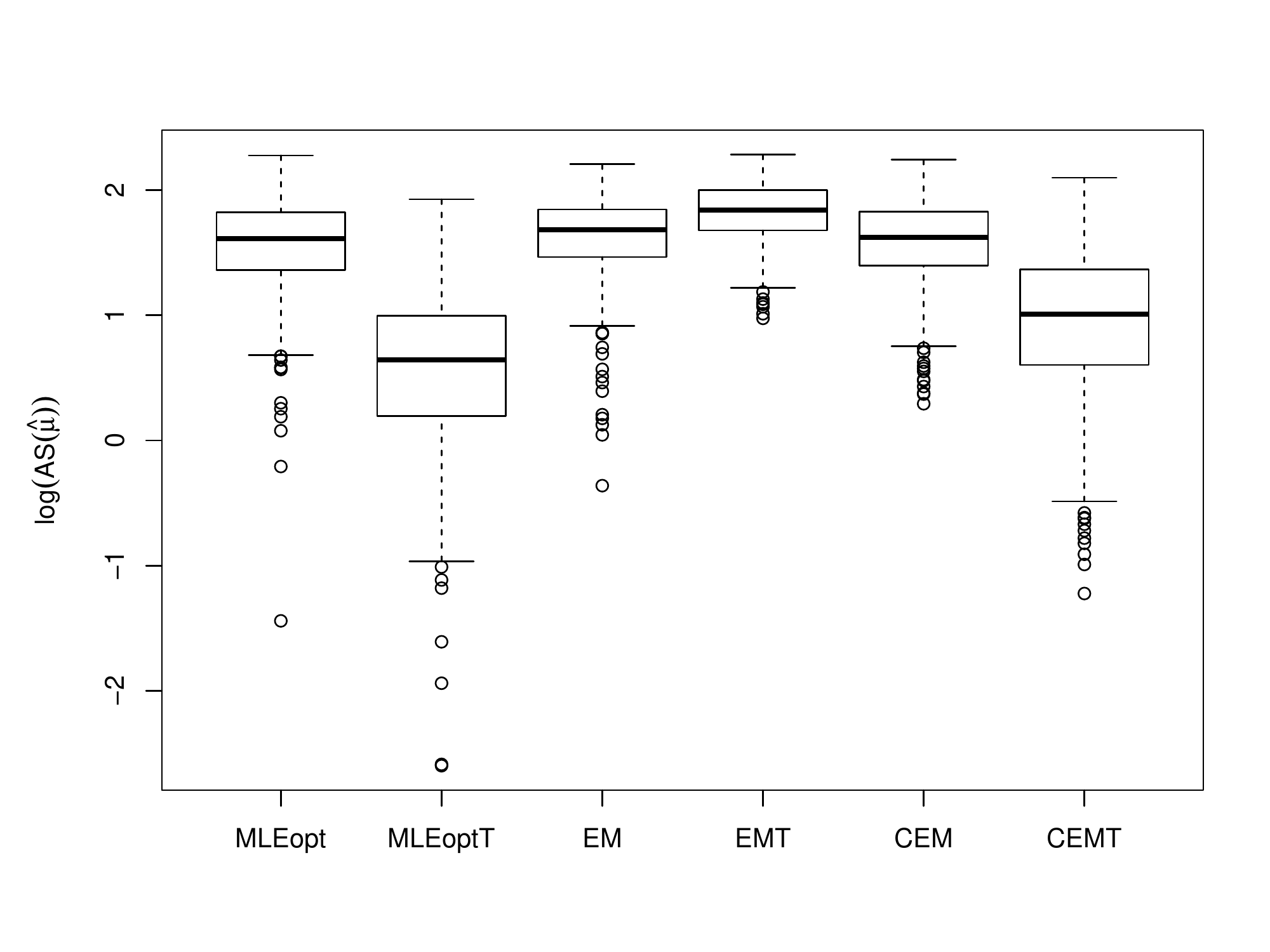} \\
\includegraphics[height=0.3\textheight]{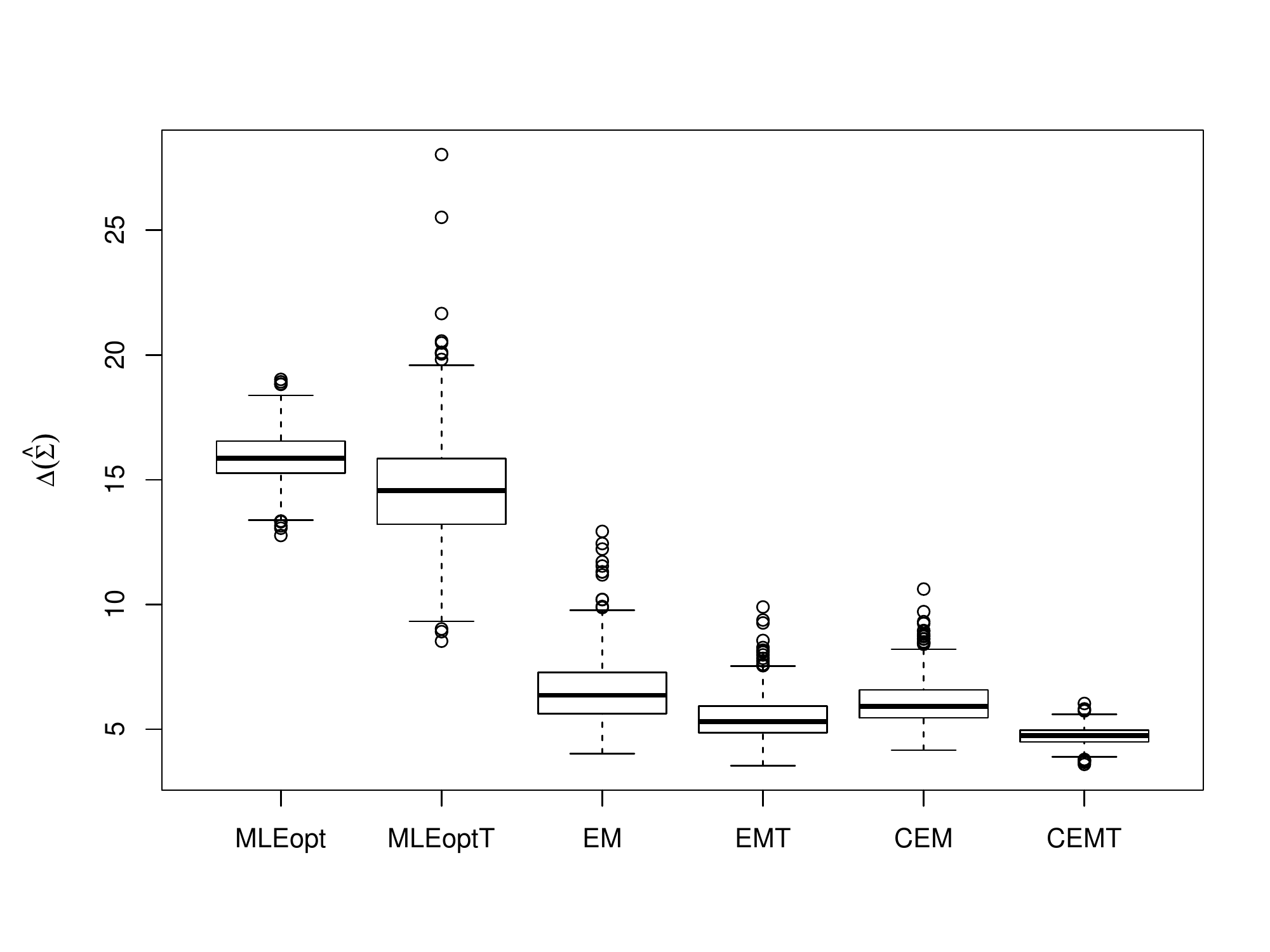} \\
\end{center}
\caption{Performance of the estimators in the case $p=5$, sample size $n=100$, $\sigma=3 \pi/2$.}
\label{fig:5:100:5}
\end{figure}

\begin{figure}
\begin{center}
\includegraphics[height=0.3\textheight]{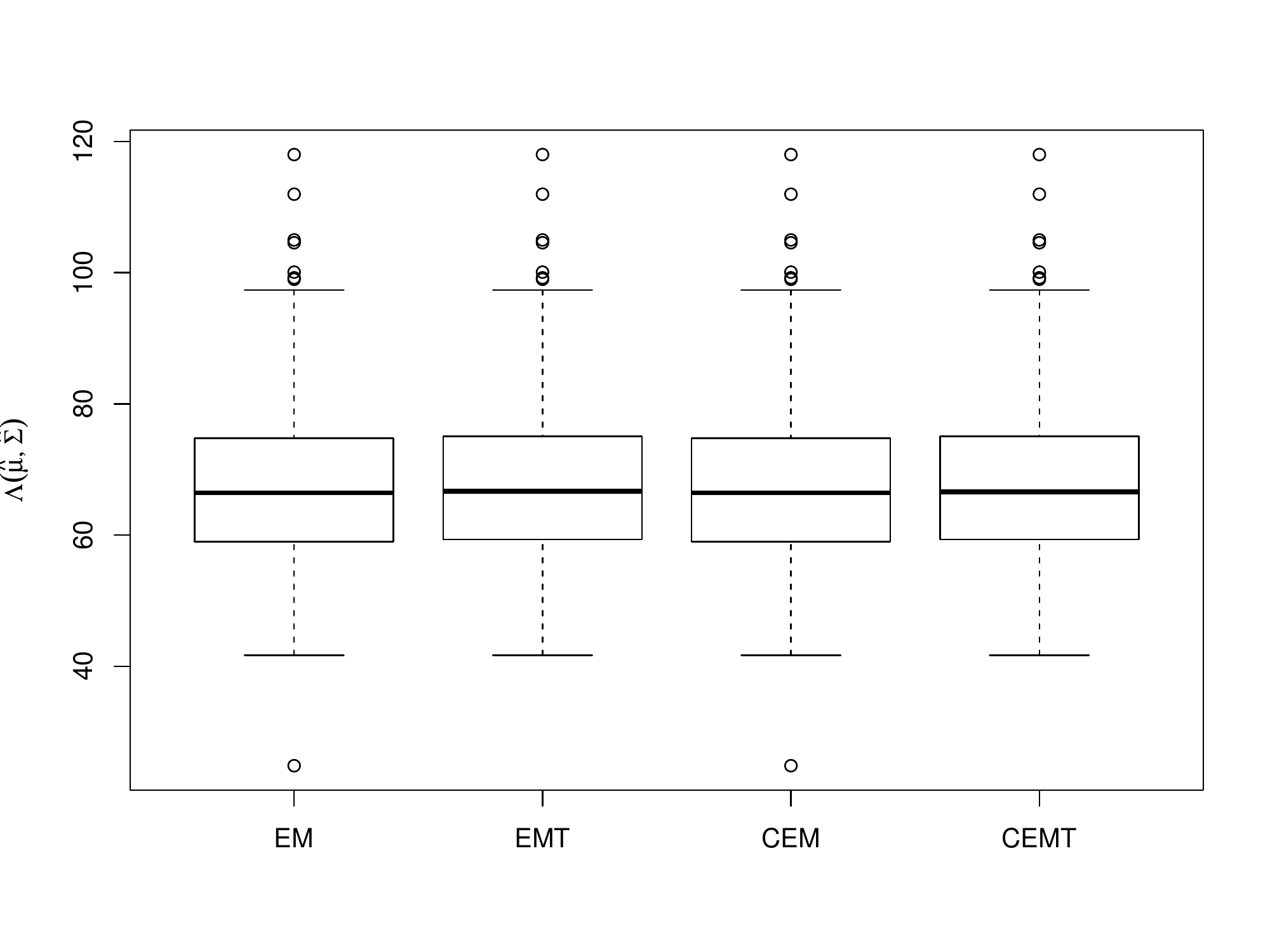} \\
\includegraphics[height=0.3\textheight]{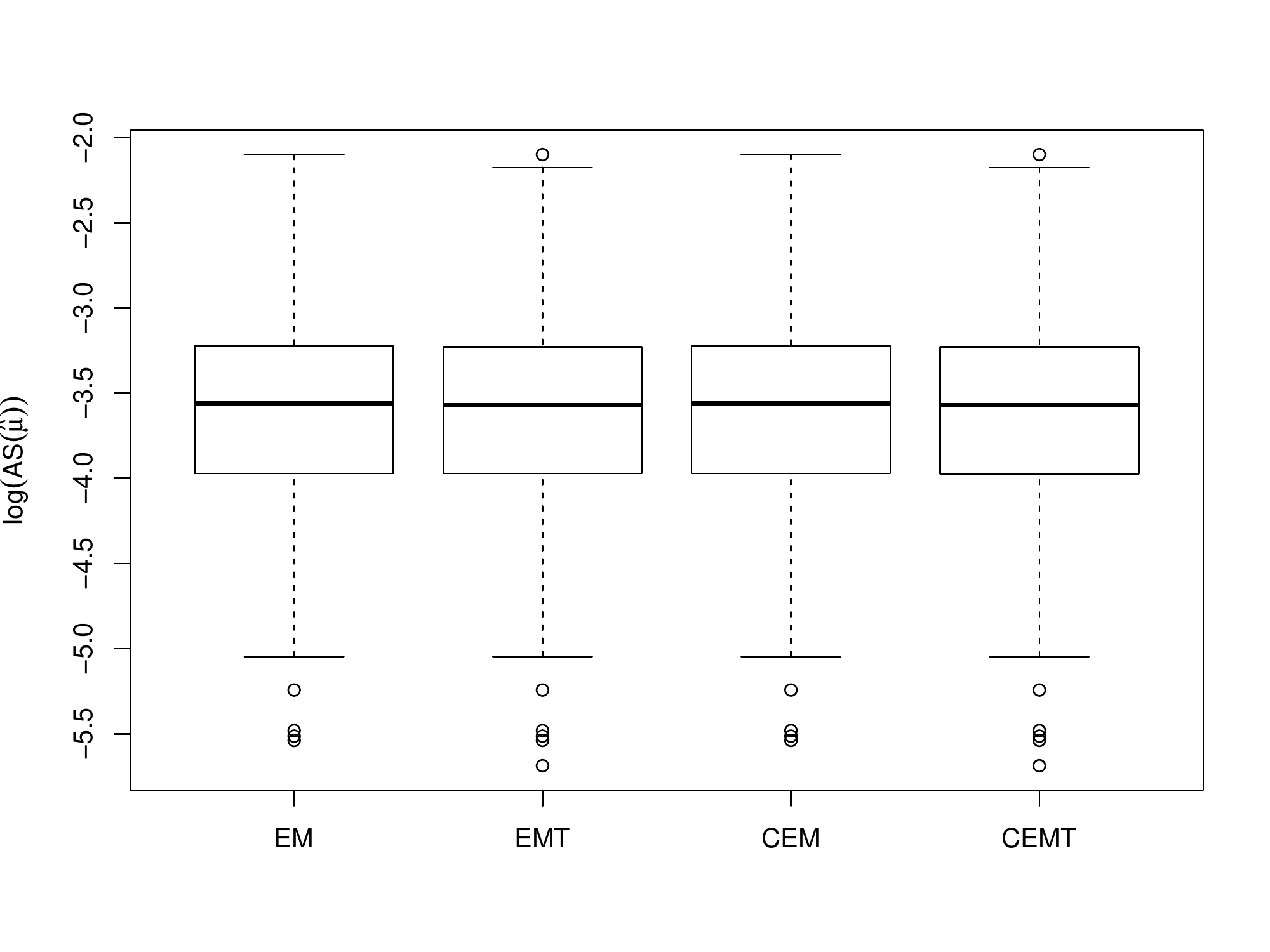} \\
\includegraphics[height=0.3\textheight]{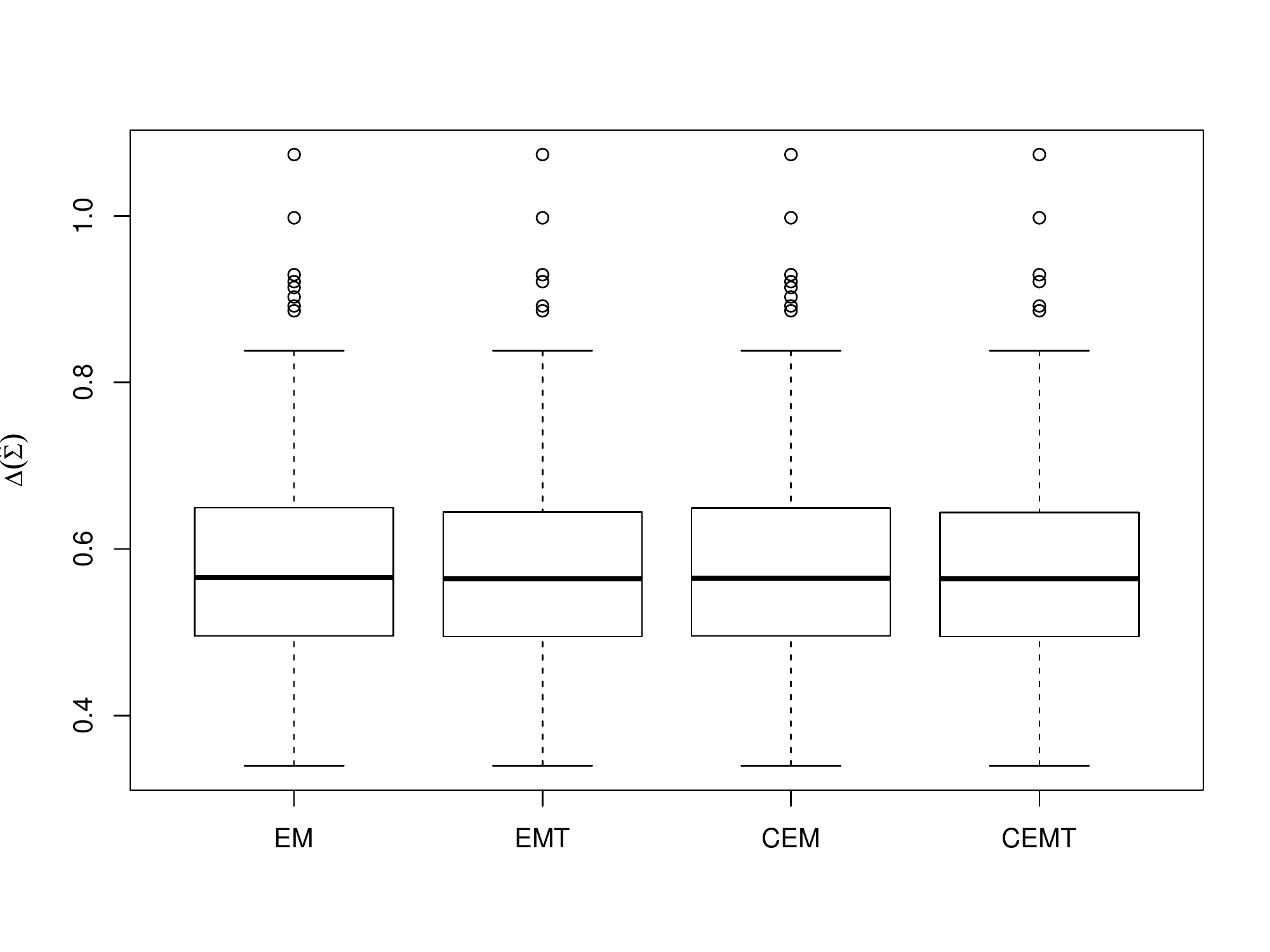} \\
\end{center}
\caption{Performance of the estimators in the case $p=10$, sample size $n=100$, $\sigma=\pi/4$.}
\label{fig:10:100:2}
\end{figure}

\begin{figure}
\begin{center}
\includegraphics[height=0.3\textheight]{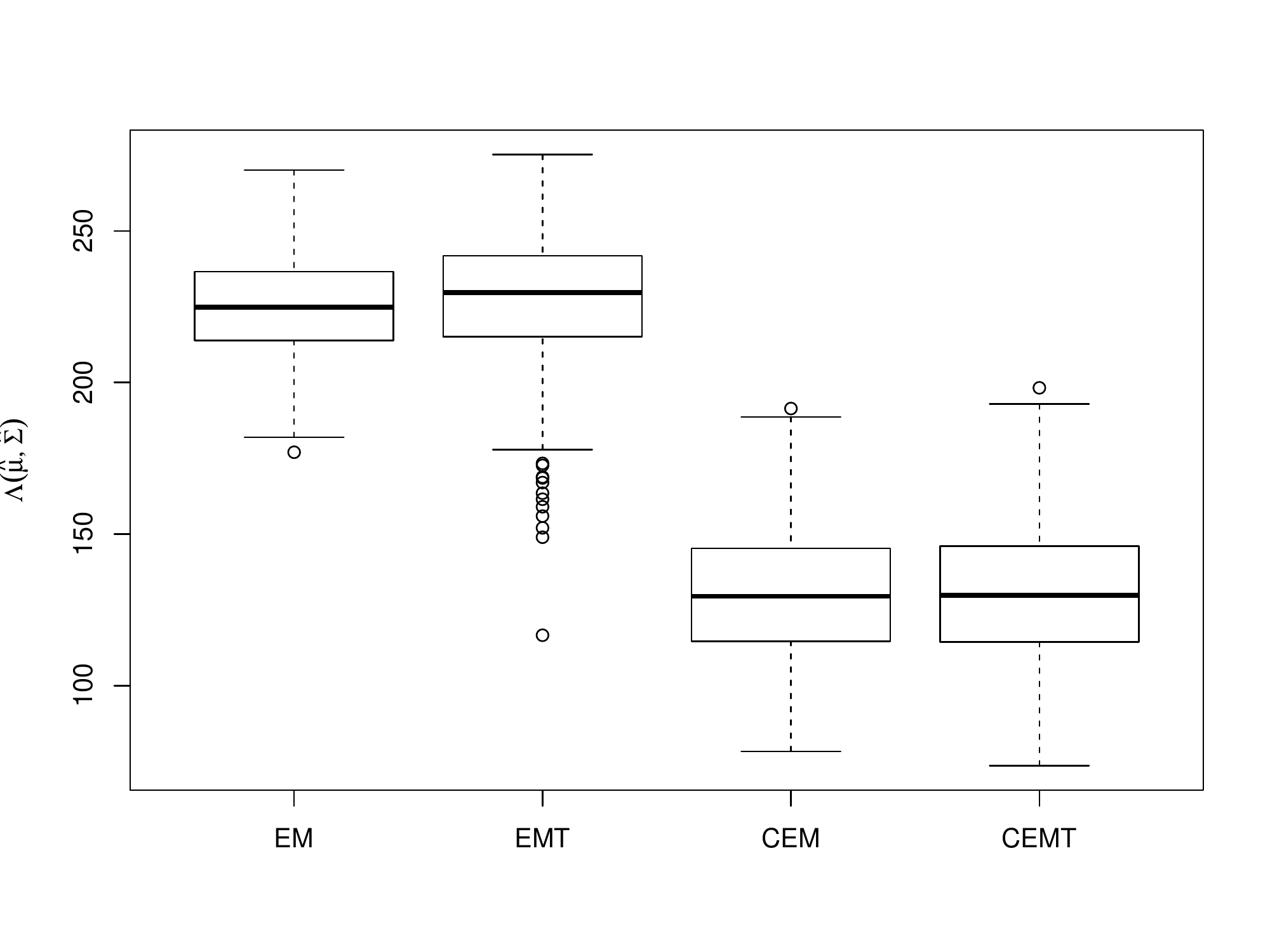} \\
\includegraphics[height=0.3\textheight]{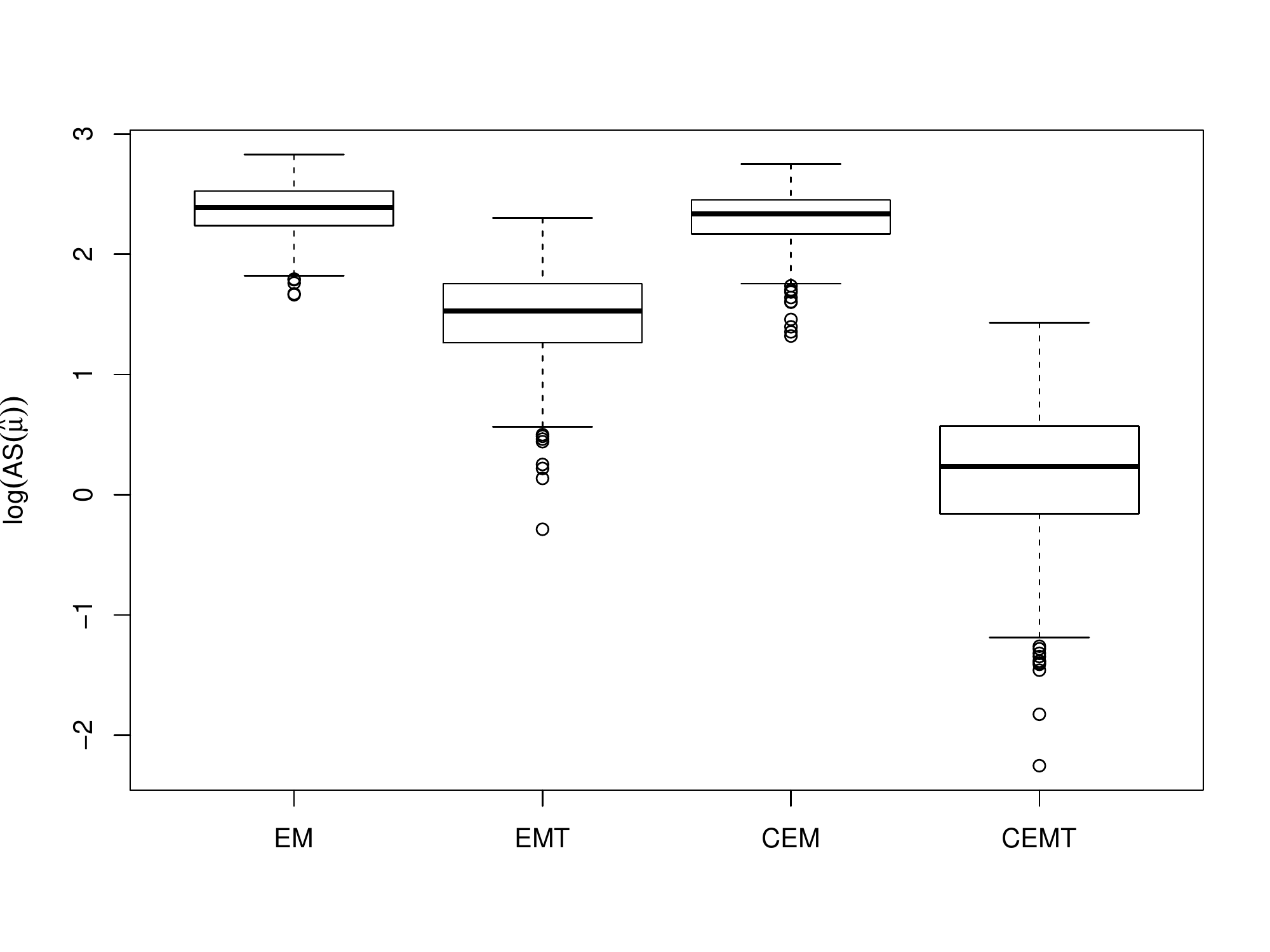} \\
\includegraphics[height=0.3\textheight]{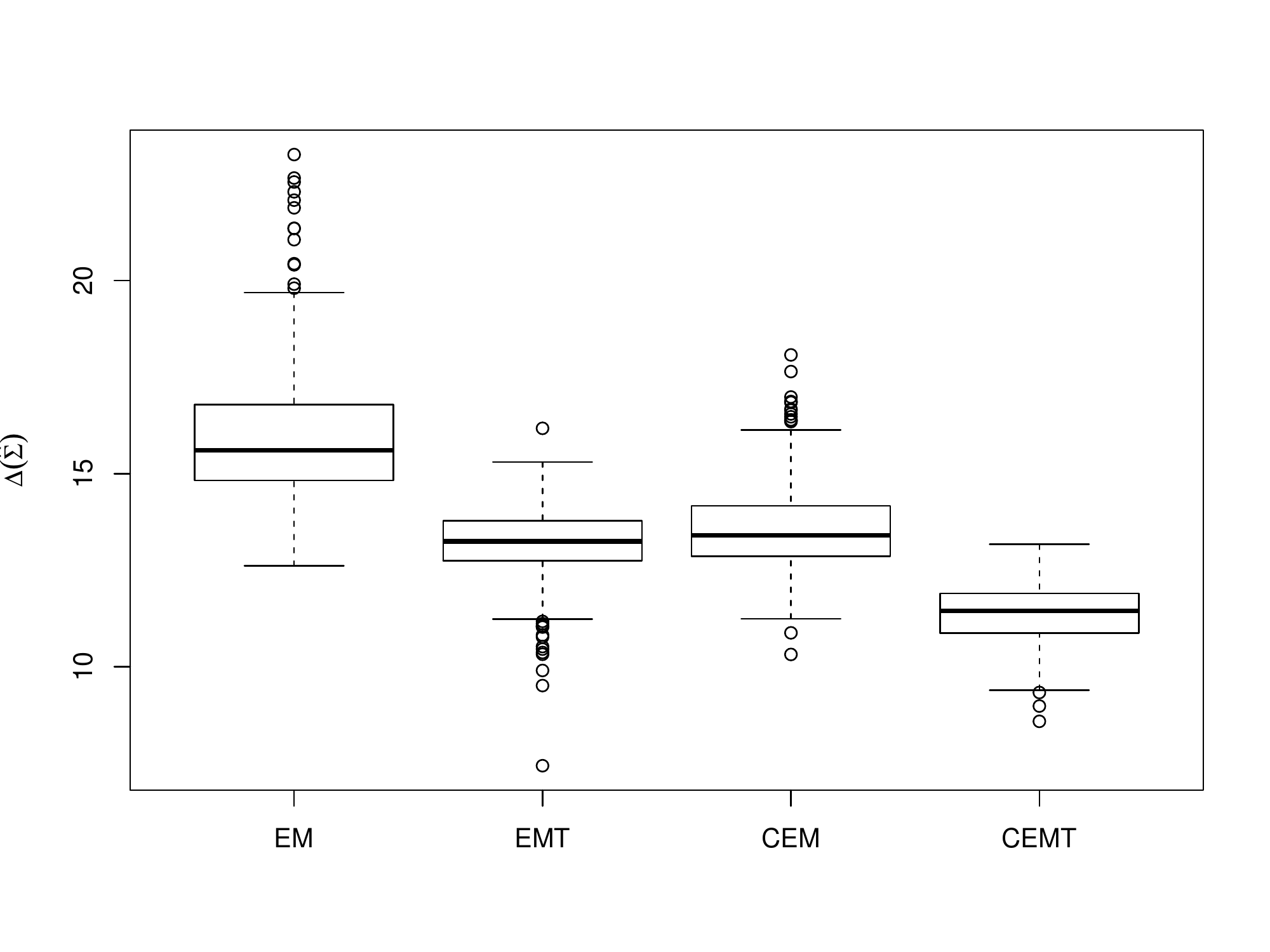} \\
\end{center}
\caption{Performance of the estimators in the case $p=10$, sample size $n=100$, $\sigma=3 \pi/2$.}
\label{fig:10:100:5}
\end{figure}

\begin{figure}
\begin{center}
\includegraphics[width=0.45\textwidth]{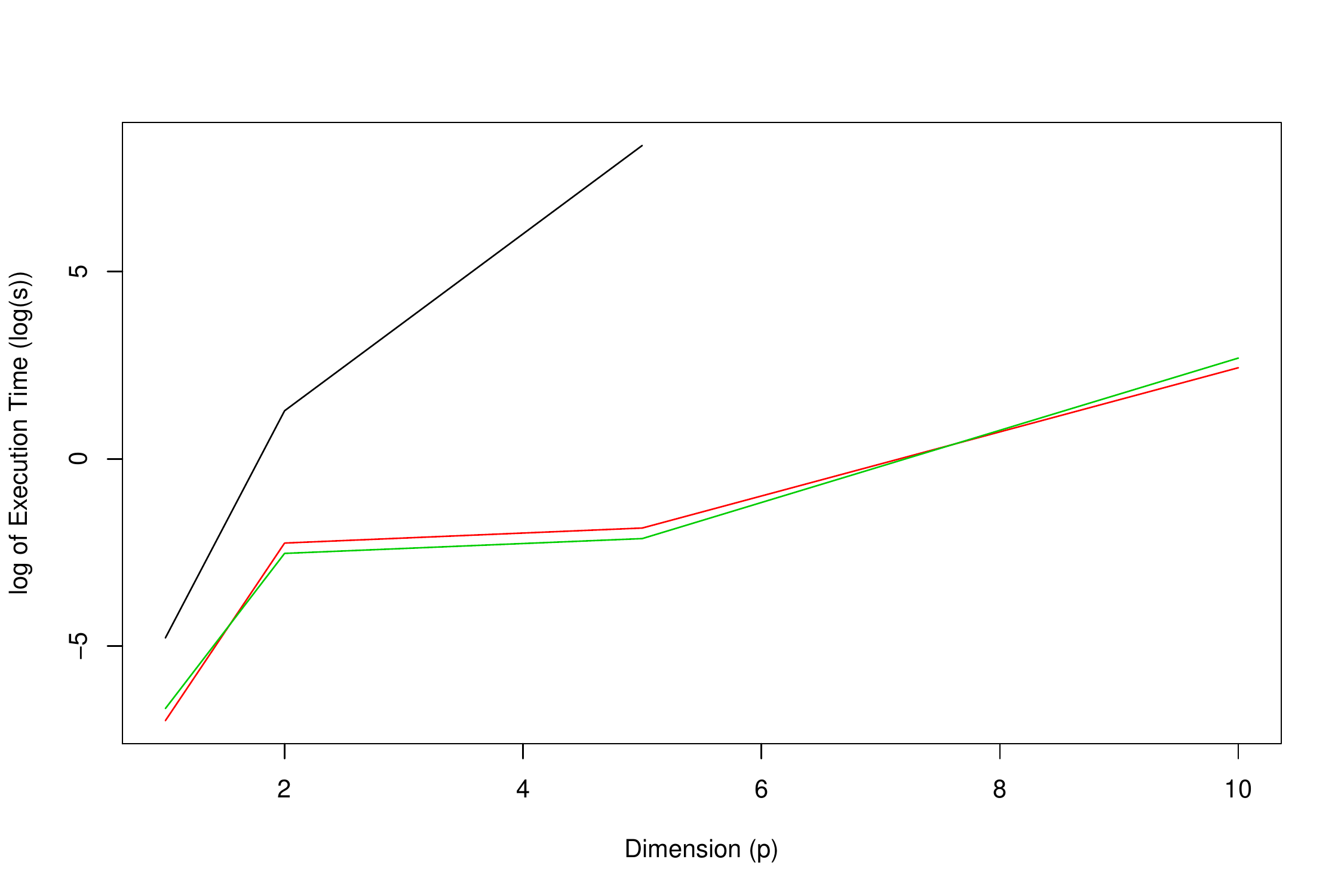}
\includegraphics[width=0.45\textwidth]{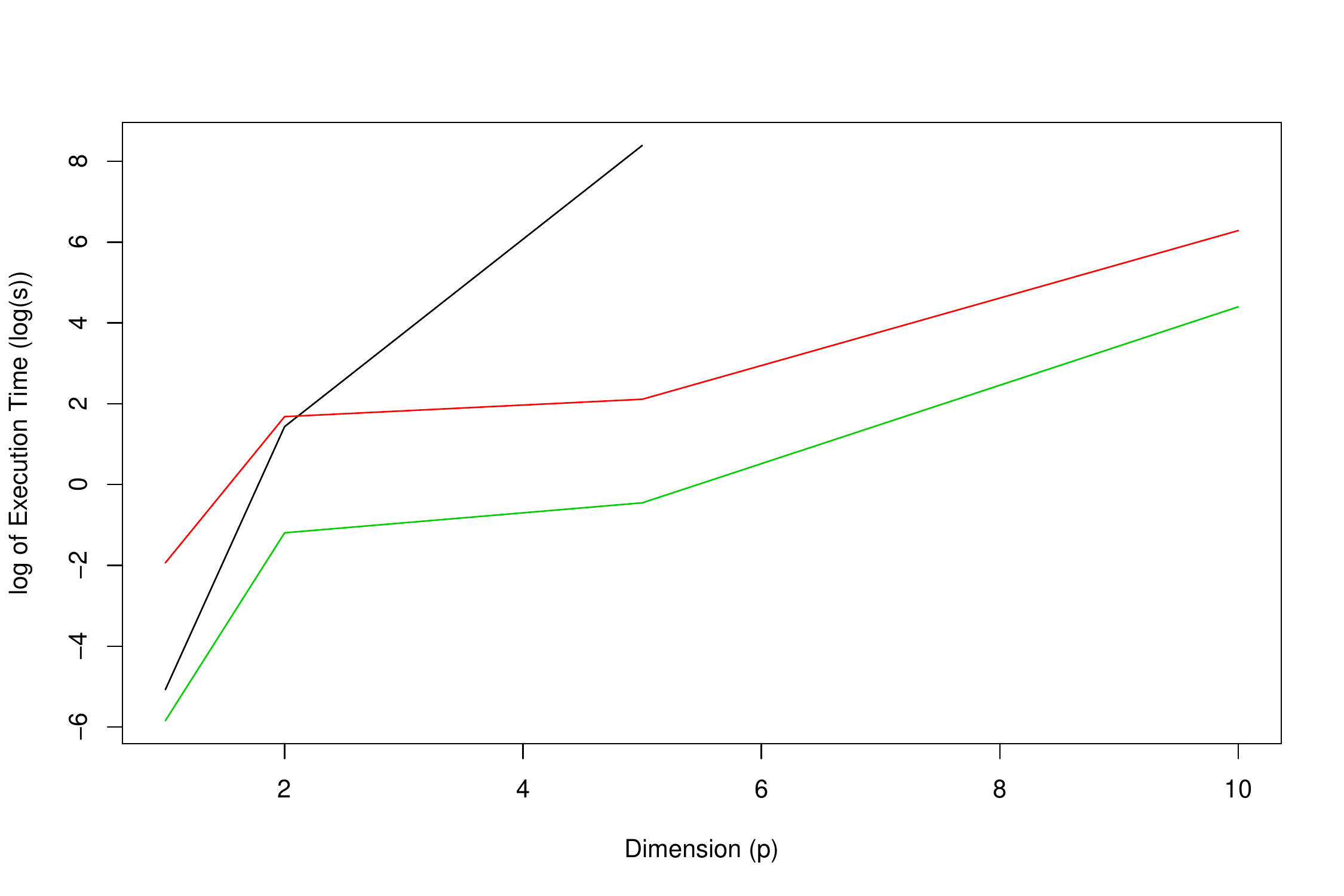} \\
\includegraphics[width=0.45\textwidth]{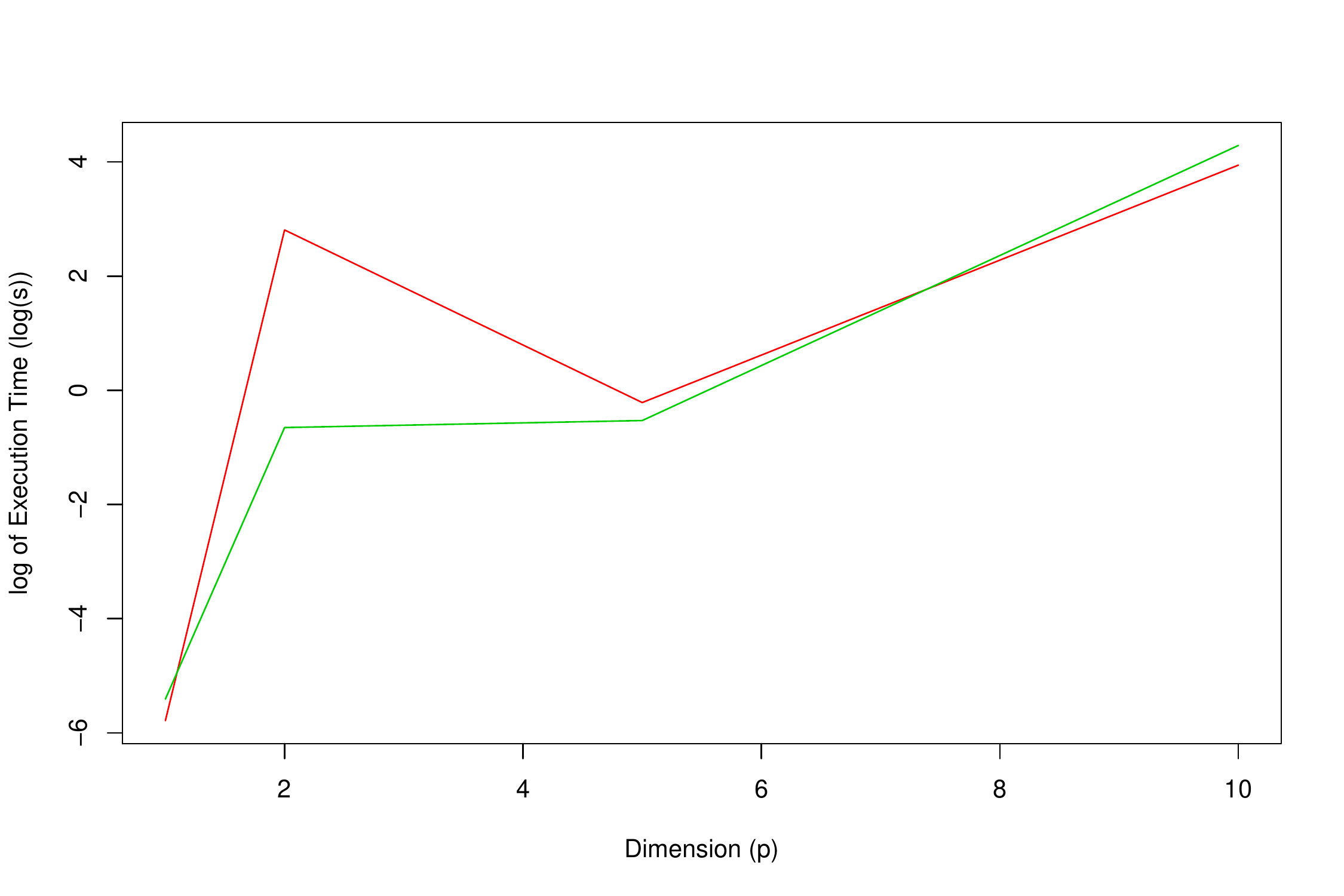}
\includegraphics[width=0.45\textwidth]{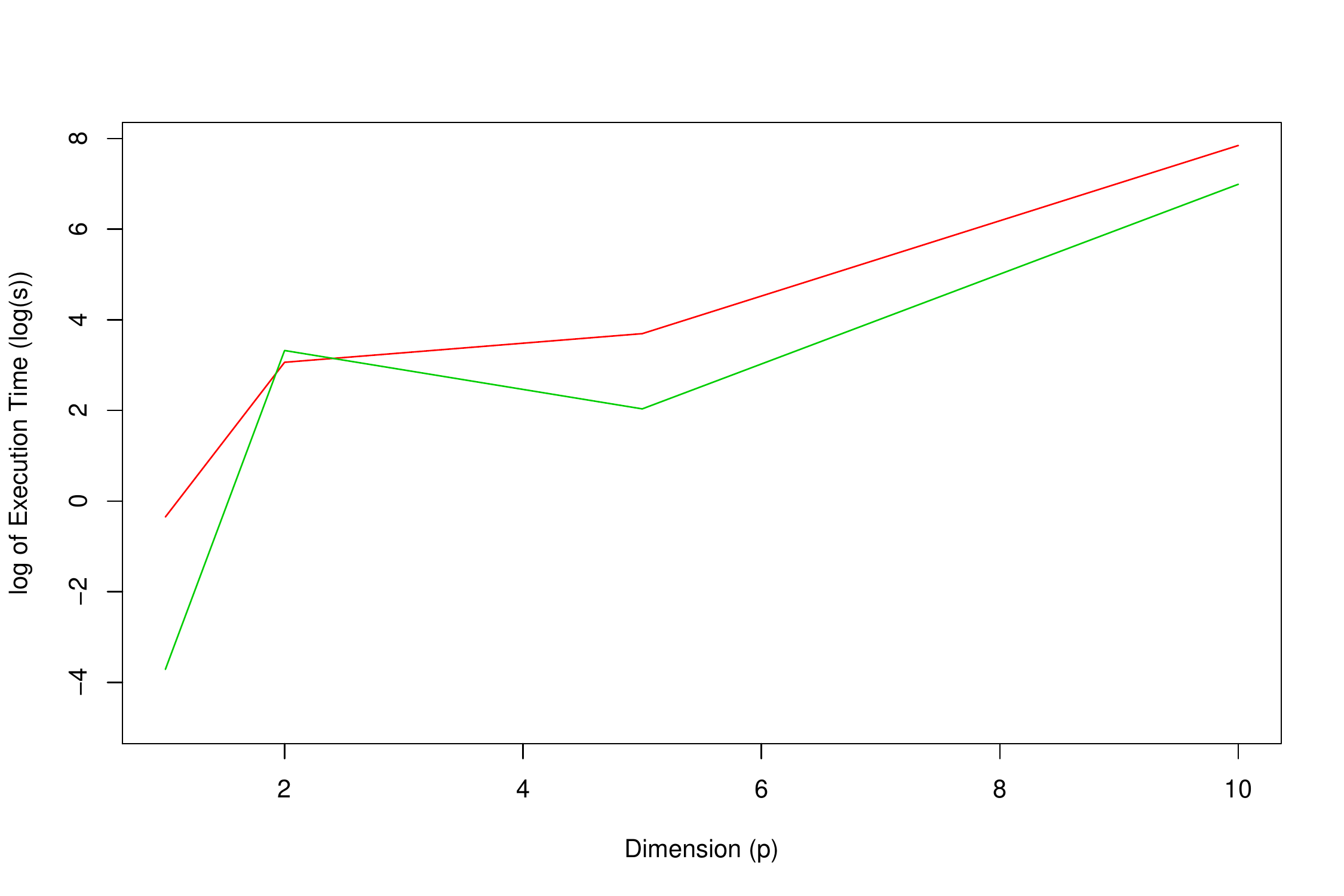}
\end{center}
\caption{Log of mean execution time for $n=100,500$ first and second rows respectively and $\sigma=\pi/8,3\pi/2$ first and second columns respectively. Black line: optim, red line: EM and green line: CEM.}
\label{fig:tempo:tutti}
\end{figure}

In Figures \ref{fig:1:100:2} -- \ref{fig:10:100:5} we report results for $n=100$ and $\sigma=\pi/4, 3 \pi/2$ for $p=1,2,5,10$ and all the considered methods, whenever their computational time were feasible. Figure \ref{fig:tempo:tutti} provides information on the execution time for $n=100,500$ and $\sigma=\pi/8,3\pi/2$. Complete results are available in Section SM-4 of the Supplementary Material.

For dimension $p=1$, small values of $\sigma$, says smaller than $\pi/2$ and all the sample sizes, all the methods perform equally well. For $\sigma=\pi/2$ and $n=10$ CEM shows an slight larger $AS$ while EM shows relatively larger $\Delta$. For larger sample size CEM and CEMT show a slight smaller $\Lambda$ and larger $\Delta$. As $\sigma$ increases further we notice that MLEoptT and EMT algorithms tend to do not move from the starting (true) values. For $\sigma \ge \pi$ MLEopt, CEM and CEMT show lower $\Lambda$, and for $n=500$ EM still provides similar results as the exact algorithm MLE. 

For $p=2$ an IRWLS algorithm is not available. As before for all sample sizes and $\sigma < \pi/2$, all the methods perform equally well. For $\sigma = \pi/2$ EM and CEM shows higher AS and $\Delta$ and a smaller $\Lambda$. This is also the case for $\sigma = \pi$, however methods starting from the true values still perform well and all similarly. For larger $\sigma$ methods show somewhat similar behavior, CEM and CEMT for larger sample size has a smaller $\Lambda$ which do not have a big impact on AS and $\Delta$ performance. 

For $p=5$ MLEopt and MLEoptT are only feasible until sample size $n=100$. As it is shown in Figure \ref{fig:tempo:tutti} average execution time in this setting is about $400s$ for $\sigma=\pi/8$ and $3000s$ for $\sigma=3\pi/2$. For $n=100$ MLEopt and MLEoptT show smaller $\Lambda$ but often also smaller AS and $\Delta$. Algorithms starting from true values show better performance for $\sigma \ge \pi/2$. For $n=500$ and $\sigma < \pi/2$ EM, EMT, CEM and CEMT show similar performance, while for larger $\sigma$ EM/EMT has larger $\Lambda$ wich has low impact on AS and $\Delta$ where methods starting from the true values still perform better.

For $p=10$ only EM and CEM are feasible. Their perfomance is similar to the case $p=5$.

\clearpage

\clearpage

\clearpage

\clearpage

\section{Conclusions}
\label{sec:conclusions}

We introduced two new algorithms based on Expectation-Maximization and Classification Expectation-Maximization methods for the estimation of the parameters in a Wrapped Normal Models for data in torus. They perform well in comparison with the direct maximization of the logarithm of the likelihood and they are still feasible in moderate high dimension data. Real examples indicate that for large dimension the introduced methods outperforms direct maximization of the log-likelihood in finding the global maximum. The methods can be easily extended to most wrapped multivariate elliptical symmetric distributions indexed by a location and scatter matrix.

\section*{Acknowledgments}

The authors thank Stephan Huckemann and Benjamin  Eltzner for preparing RNA data set.

\bibliographystyle{elsarticle-harv}

\end{document}